\documentstyle[12pt]{article}
\begin{document}
\def \inbar{\vrule height1.5ex width.4pt depth0pt}
\def \C{\relax\hbox{\kern.25em$\inbar\kern-.3em{\rm C}$}}
\def \R{\relax{\rm I\kern-.18em R}}
\newcommand{\Z}{\ Z \hspace{-.08in}Z}
\newcommand{\be}{\begin{equation}}
\newcommand{\ee}{\end{equation}}
\newcommand{\bea}{\begin{eqnarray}}
\newcommand{\eea}{\end{eqnarray}}
\newcommand{\p}{\psi}
\renewcommand{\l}{\epsilon}
\newcommand{\f}{\phi}
\newcommand{\g}{\gamma}
\newcommand{\G}{\Gamma}
\newcommand{\e}{\eta}
\newcommand{\z}{\pi}
\newcommand{\m}{\mu}
\newcommand{\n}{\nu}
\newcommand{\s}{\sigma}
\renewcommand{\t}{\tau}
\renewcommand{\a}{\alpha}
\renewcommand{\b}{\beta}
\newcommand{\k}{\kappa}
\renewcommand{\d}{\delta}
\newcommand{\th}{\theta}
\renewcommand{\pt}{\frac{\partial}{\partial t}}
\newcommand{\ppt}{\frac{\partial^{2}}{\partial t^{2}}}
\newcommand{\nn}{\nonumber}
\renewcommand{\ll}{\left[ }
\newcommand{\rr}{\right] }
\newcommand{\kt}{\rangle}
\newcommand{\br}{\langle}
\newcommand{\fs}{\small}
\newcommand{\so}{S_{0}}
\newcommand{\I}{\mbox{1}_{m\times m}}
\newcommand{\In}{\mbox{1}_{n\times n}}
\newcommand{\xo}{x_{0}}
\newcommand{\po}{\psi_{0}}
\newcommand{\pss}{\frac{\partial}{\partial s}}
\newcommand{\tcuf}{\tilde{\cal F}}

\title{SUPERSYMMETRY AND THE ATIYAH-SINGER INDEX THEOREM I:
Peierls Brackets, Green's Functions, and a Supersymmetric Proof of
the Index Theorem}
\author{Ali Mostafazadeh \\ \\ Center for Relativity \\ The University
of Texas at Austin \\ Austin, Texas 78712, USA}
\maketitle
\baselineskip=18pt
\begin{abstract}
The Peierls bracket quantization scheme is applied to the supersymmetric
system corresponding to the twisted spin index theorem.
A detailed study of the quantum system is presented, and the Feynman
propagator is exactly computed. The Green's function methods provide
a direct derivation of the index formula.
\end{abstract}
\newpage

\section{Introduction}
The Peierls bracket quantization is explained and applied to many
interesting examples in \cite{bd1}. In particular, it is used
to quantize a supersymmetric system described by the Lagrangian:
\be
L=\mbox{\fs$\frac{1}{2}$}g_{\mu \nu}\dot{x}^{\mu}\dot{x}^{\nu}+
\mbox{\fs$\frac{i}{2}$}g_{\mu \nu}
\psi_{\alpha}^{\mu}\frac{D}{dt}(\psi_{\alpha}^{\nu})
+\mbox{\fs$\frac{1}{8}$}
R_{\mu \nu \sigma \tau}\psi_{\alpha}^{\mu}\psi_{\alpha}^{\nu}
\psi_{\beta}^{\sigma}\psi_{\beta}^{\tau}
\label{eq1}
\ee
\[ \alpha , \beta = 1 , 2 \hspace{.5in} \mu ,\nu ,\sigma ,\tau =1,...,m.\]
The configuration space of (\ref{eq1}) is an $(m,2m)$-dimensional
supermanifold \cite{bd1} with $x$ and $\psi$ denoting its bosonic
(commuting) and fermionic (anticommuting) coordinates, respectively.
The quantization of (\ref{eq1}) gives rise to a rigorous supersymmetric
proof of the Gauss-Bonnet(-Chern-Avez) theorem, \cite[Part I, p. 395]{cd1}.

The Gauss-Bonnet theorem is the best known example of an index theorem. It
is also called the index theorem for the deRham complex \cite{egh}. The
idea of using supersymmetric quantum mechanics to give proofs of
index theorems was originally suggested by Witten who discovered
the relation between the two subjects in his study of supersymmetry
breaking, \cite{witten}. Windey \cite{windey} and Alvarez-Gaume
\cite{ag1,ag2,ag3} provided the first supersymmetric proofs of the
index theorem. Few years later, WKB methods were applied to
give an alternative approach to the path integral evaluation of the
index by Ma\~{n}es and Zumino, \cite{manes-zumino}. The same was
achieved in a concise paper by Goodman \cite{goodman} who used
the mode expansion techniques. Among other remarkable works on this
subject are a difficult paper by Getzler \cite{getzler} and a
superspace approach by Friedan and Windey \cite{friedan-windey}.

The present paper uses the basic Green's function methods discussed
in \cite{bd1} to arrive at a proof of the index theorem. The
strategy is the same as the earlier supersymmetric proofs, but
the methods differ appreciably. In particular, the Feynman
propagators are computed exactly for the first time.

In Section 2, the Atiyah-Singer index theorem and its relation to
supersymmetric quantum mechanics are reviewed. Section 3 starts with
a brief description of the Peierls bracket quantization. The
superclassical system corresponding to the twisted spin complex is
then introduced and its quantization is performed. The quantum
mechanical Hamiltonian involves a scalar curvature factor.
Section 4 includes a discussion of the quantum systems associated
with the spin and twisted spin complexes. In particular, it is shown
that the quantization of the supersymmetric charge leads to its
identification with the corresponding Dirac operator. In Section 5,
the path integral representation of the index is presented and the
Green's function methods are discussed. Sections 6 and 7 are devoted
to the proofs of the spin and twisted spin index theorems, respectively.
Section 8 includes the final remarks.

Throughout this paper $\hbar$ will be set to 1, except in Section 3.

\section{The Atiyah-Singer Index Theorem and Supersymmetric Quantum
Mechanics}

The Atiyah-Singer index theorem is one of the most substantial
achievements of modern mathematics. It has been a developing subject
since its original proof by Atiyah and Singer, \cite{atiyah-singer}.
A clear exposition of the index theorem is given in the classic paper
of Atiyah, Bott, and Patodi, \cite{abp}. For the historical origins
of the index theory, see \cite{atiyah}. Several mathematical proofs of
the index theorem and its variations and generalizations are available
in the literature, \cite{palias,shanahan,gilkey,booss,berline}. A
common feature of all these proofs is the use of K-theory, \cite{atiyah-k}.
In particular, the original cobordism proof \cite{atiyah-singer}, the
celebrated heat kernal proof \cite{abp,gilkey}, and the supersymmetric
proofs, \cite{windey,ag3}, of the index theorem are based on a result
of K-theory which reduces the proof of the ``general'' index theorem
to the special case of twisted signature or alternatively twisted spin
index theorems, \cite{abp,shanahan,gilkey,booss}.

In ``general'', one has an elliptic differential operator,
\be
D: {\cal C}^{\infty}(V_{+})\longrightarrow {\cal C}^{\infty}(V_{-})
\label{eq2}
\ee
between the spaces of sections of two hermitian vector bundles
\footnote{A hermitian vector bundle is a complex vector bundle
which is endowed with a hermitian metric and a compatible
connection \cite{abp}.}
$V_{+}$ and $V_{-}$ over a closed \footnote{compact, without boundary}
Riemannian manifold, $M$. Alternatively one can speak of the  short
elliptic complex:
\[ 0 \longrightarrow {\cal C}^{\infty}(V_{+}) \stackrel{D}{\longrightarrow}
{\cal C}^{\infty}(V_{-}) \longrightarrow 0 . \]
The index theorem provides a closed formula for the analytic index of
$D$. The latter is defined by
\be
index(D) := dim[ker(D)]-dim[coker(D)]\; .
\label{eq3}
\ee
The symbols dim, ker, and coker are abbreviations of {\em dimension,
kernel}, and {\em cokernel}, respectively. Indices of elliptic operators
are of great interest in mathematics and physics because they are
topological invariants.

The ``general'' index formula computes the index as an integral
involving some characteristic classes associated with the vector
bundles $V_{\pm}$, the base manifold $M$, and the operator
$D$. The twisted (or generalized) spin index theorem is a special case.

Let $M$ be a closed $m=2l$-dimensional spin manifold, \cite{egh}. Let
${\cal S}={\cal S}^{+}\oplus {\cal S}^{-}$ denote the spin bundle
of M. The sections of ${\cal S}^{\pm}$ are called the $\pm$ chirality
spinors. Let $\not\!\! D:{\cal C}^{\infty}({\cal S}) \rightarrow {\cal C}^
{\infty}({\cal S})$ be the Dirac operator \cite{shanahan,gilkey,egh,nash},
and $\not\!\partial :=\not\!\!D \mid_{{\cal C}^{\infty}({\cal S}^{+})}$. The
short complex
\be
0 \longrightarrow {\cal C}^{\infty}({\cal S}^{+}) \stackrel{\not
\hspace{.4mm}\partial}{\longrightarrow}
{\cal C}^{\infty}({\cal S}^{-}) \longrightarrow 0
\label{eq3.1}
\ee
is called the spin complex.

\newtheorem{thm}{Theorem}
\begin{thm}
Let $\not\!\partial$ be as in (\ref{eq3.1}), and  define the so-called
$\hat{A}$-genus density:
\be
\hat{A}(M) := \prod_{i=1}^{l} \left[ \frac{\frac{\Omega_{i}}{4\pi}}{\sinh
(\frac{\Omega_{i}}{4\pi})} \right].
\label{eq3.2}
\ee \newcommand{\aroof}{\int_{M}\left[ \hat{A}(M) \right]_{top}}
where, $\Omega_{i}$ are the 2-forms defined by block-diagonalizing the
curvature 2-form \footnote{The block-diagonalization is always possible since
$\Omega$ is antisymmetric.}:
\be
\Omega :=(\mbox{\fs$\frac{1}{2}$}
R_{\mu \nu \gamma \lambda} dx^{\gamma}\wedge dx^{\lambda})
=: diag \left( \left[ \begin{array}{cc}
0 & \Omega_{i} \\
-\Omega_{i} & 0
\end{array} \right] : i=1 \cdots l \right).
\label{eq5}
\ee
Then,
\be
index(\not\!\partial) = \aroof .
\label{eq4}
\ee
\end{thm}
In (\ref{eq4}), ``top'' means that the highest rank form in the power series
expansion of (\ref{eq3.2}) is integrated.

Let $V$ be a hermitian vector bundle with fiber dimension $n$, base
manifold M, and connection 1-form $A$. Then, the operator:
\[\not\!\! D_{V} :{\cal C}^{\infty}({\cal S}^{\pm})\otimes {\cal C}^{\infty}(V)
\longrightarrow {\cal C}^{\infty}({\cal S}^{\mp})\otimes {\cal C}^{
\infty}(V) \]
defined by
\be
\not\!\! D_{V}(\Psi \otimes v) :=\not\!\! D(\Psi)\otimes v + (-1)^{\Psi}\Psi
\otimes D_{A}(v),
\label{eq6}
\ee
is an elliptic operator, called the {\em twisted Dirac } operator.
In (\ref{eq6}), \\ $\Psi \in{\cal C}^{\infty}({\cal S}^{\pm})
\Leftrightarrow (-1)^{\Psi}=\pm 1$ , $v\in {\cal C}^{\infty}(V)$ ,
and $D_{A}$ is the covariant derivative operator defined by $A$,
\cite{gilkey}. Also define
\be
\not\!\partial_{V} :=\not\!\! D_{V}\mid_{{\cal C}^{\infty}({\cal S}^{+})
\otimes{\cal C}^{\infty}(V)}.
\label{eq6.1}
\ee
The twisted spin complex is the following short complex:
\be
0 \longrightarrow {\cal C}^{\infty}({\cal S}^{+}\otimes V) \stackrel{
\not\hspace{.4mm}\partial_{V}}{\longrightarrow} {\cal C}^{\infty}({\cal S}^{-}
\otimes V)\longrightarrow 0.
\label{eq7}
\ee
\begin{thm}
Let $\not\!\partial_{V}$ be as in (\ref{eq6.1}), and
\be
ch(V):=tr\left[ \exp\frac{iF}{2\pi} \right]
\label{eq9}
\ee
be the Chern character of $V$. Then,
\be
index(\not\!\partial_{V})= \int_{M}\left[ ch(V).\hat{A}(M) \right]_{top}.
\label{eq8}
\ee
\end{thm}
In (\ref{eq9}), $F$ is the curvature 2-form of the connection 1-form
$A$ (written in a basis of the structure Lie algebra of $V$):
\be
F:=\left(\mbox{\fs$\frac{1}{2}$} F_{\lambda \gamma}^{ab}
\, dx^{\lambda}\wedge dx^{\gamma}
\right)\; .
\label{eq9.1}
\ee

Throughout this paper, the Greek indices refer to the coordinates
of $M$, they run through $1,...,m=dim(M)$, and the indices from
the beginning of the Latin alphabet refer to the fibre coordinates
of $V$ and, hence, run through $1,...,n$.

The relevance of the index theory to supersymmetry has been discussed
in almost every article written in this subject in the past ten years.
The idea is to realize the parallelism between the constructions
involved in the index theory, (\ref{eq2}), and the supersymmetric
quantum mechanics. In the latter, the Hilbert (Fock) space is the
direct sum of the spaces of the bosonic and the fermionic state vectors.
These correspond to the spaces of sections of $V_{\pm}$ in (\ref{eq2}).
Moreover, the supersymmetric charge $\cal Q$ plays the role of the
elliptic operator $D$, and the Hamiltonian $H$ is the analog of the
Laplacian $\Delta$ of $D$. One has:
\[ \Delta :=\{ D,D^{\dagger}\} , \]
and
\be
H=\frac{1}{2}\{ {\cal Q},{\cal Q}^{\dagger}\}.
\label{eq10}
\ee
Here, ``$\dagger$'' denotes the adjoint of the corresponding operator.
(\ref{eq10}) is also known as the superalgebra condition. One can also
define selfadjoint supersymmetric charges \footnote{
$\alpha =1,...,2N$, where N is the number of nonselfadjoint
charges (type N-SUSY).}, $Q_{\alpha}$ \cite{susy},
in terms of which (~\ref{eq10}) becomes:
\be
H=Q_{\alpha}^{2}\; , \; \forall \a \; .
\label{eq10.1}
\ee
The next step is to recall
\[ coker(D) = ker(D^{\dagger})\; , \]
\[ ker(\Delta) = ker(D) \oplus ker(D^{\dagger}) \]
and use (\ref{eq3}), and (\ref{eq9}) to define:
\[ index_{W} := n_{b,0}-n_{f,0}. \]
Here``$W$'' refers to Witten \cite{witten} and $n_{b,0}$ and $n_{f,0}$
denote the number of the zero-energy bosonic and fermionic
eigenstates of $H$. Realizing that due to supersymmetry any excited
energy eigenstate has a superpartner, one has the following set
of equalities:
\bea
index_{W} & = & n_{b}-n_{f}  \nonumber \\
& = & tr[(-1)^{f}]  \label{eq11} \\
& = & tr[(-1)^{f}e^{-i\beta H}] \nonumber \\
& =: & str[e^{-i\beta H}]. \nn
\eea
In (\ref{eq11}), $n_{b}$ and $n_{f}$ denote the number of the bosonic
and fermionic energy eigenstates, and $f$ is the fermion number
operator. See \cite{witten,susy,ag3} for a more detailed discussion
of (\ref{eq11}).

The appearance of $e^{-i\beta H}$ in (\ref{eq11}) is quite interesting.
It had, however, been noticed long before supersymmetry was introduced
in physics. The heat kernel proof of the index theorem is essentially
based on (\ref{eq11}), \cite{abp,booss,gilkey} \footnote{In the heat
kernel proof, one computes the $index$ using the formula:
\[ index = tr[ e^{-\b \Delta_{+}}] - tr[ e^{-\b \Delta_{-}}] \; ,\]
where $\Delta_{+} := D^{\dagger}D$ and $\Delta_{-} := D\,
D^{\dagger}$.}. The major advantage of
the supersymmetric proofs is that one can compute $index_{W}$ using
its path integral representation. In particular, since $index_{W}$
is independent of $\beta$, the WKB approximation, i.e. the first
term in the loop expansion, yields the index immediately.

\section{The Superclassical System and Its  Quantization}
Consider a superclassical system described by the action functional
$S=S[\Phi ]$, with $\Phi = (\Phi^{i})$ denoting the coordinate (field)
variables. The dynamical equations are given by $\delta S = 0$, i.e.
\be
S,_{i} := S[\Phi ]\frac{\stackrel{\leftarrow}{\delta}}{\delta
\Phi^{i}(t)} = 0.
\label{eq12}
\ee
Throughout this paper the condensed notation of \cite{bd1}
is used where appropriate. The following example demonstrates most
of the conventions:
\[ _{i,}S_{,j'}G^{j'k''}\equiv \int dt'\left(
\frac{\stackrel{\rightarrow}{\delta}}{\delta \Phi^{i}(t)} S[\Phi ]
\frac{\stackrel{\leftarrow}{\delta}}{\delta \Phi^{j}(t')} \right)
G^{jk}(t',t'') . \]
In general, the second functional derivatives of $S$ ``are'' second
order differential operators \footnote{These are called the Jacobi
operators.}, e.g. see (\ref{eq24}) below. Let $G^{\pm i'k''}$
denote the corresponding advanced and retarded Green's functions:
\be
_{i,}S_{,j'}G^{\pm j'k''} := - \delta(t-t'')\delta_{i}^{k}
\label{eq13}
\ee
where $\delta(t-t'')$ and $\delta_{i}^{k}$ are the Dirac and Kronecker
delta functions, respectively. Furthermore, define:
\be
\tilde{G}^{jk'} := G^{+jk'} - G^{-jk'}\; .
\label{eq14}
\ee
Then, the Peierls bracket \cite{peierls,bd1} of
any two scalar fields, ${\cal A}$  and ${\cal B}$, of $\Phi$ is defined
by:
\be
({\cal A},{\cal B}') := {\cal A}_{,i}\hspace{1mm}\tilde{G}^{ij'}\hspace{.1mm}
_{j',}{\cal B}.
\label{eq14.9}
\ee
In particular, one has
\be
(\Phi^{i},\Phi^{j'}) := \tilde{G}^{ij'}.
\label{eq15}
\ee

The Green's functions $G^{\pm ij'}$ satisfy the following reciprocity
relation \cite{bd1},
\be
G^{-ij'} = (-1)^{ij'} G^{+j'i}.
\label{eq15.1}
\ee
In (\ref{eq15.1}), the indices $i$ and $j'$ in $(-1)^{ij'}$ are either
0 or 1 depending on whether $\Phi^{i}$($\Phi^{j'}$) is a bosonic
or fermionic variable, respectively.

The quantization is performed by promoting the superclassical quantities
to the operators acting on a Hilbert space and forming a superalgebra
defined by the following supercommutator:
\be
[{\cal A},{\cal B}]_{\rm super} := i \hbar ({\cal A},{\cal B})\; ,
\label{eq16}
\ee
where the right hand side is defined up to factor ordering.

The superclassical system of interest is represented by the
Lagrangian \cite{windey,ag3}:
\bea
L&=&\left[ \mbox{\fs$\frac{1}{2}$}g_{\mu \nu}(x) \dot{x}^{\mu}\dot{x}^{\nu}+
\mbox{\fs $\frac{i}{2}$}
g_{\lambda \gamma}(x)\psi^{\lambda}\frac{D}{dt}\psi^{\gamma}
\right]_{1} + \label{eq17} \\
&&+\kappa \left[ i\eta^{a*}\left( \dot{\eta}^{a}+\dot{x}^{\sigma}
A_{\sigma}^{ab}(x)\eta^{b}\right) +\mbox{\fs$\frac{1}{2}$}
F_{\lambda \gamma}^{ab}(x) \psi^{\lambda}\psi^{\gamma}\eta^{a*}\eta^{b}
\right]_{2} + \nonumber \\
&& \hspace{.5cm}+\left[ \mbox{\fs$\frac{\alpha}{\beta}$}
\eta^{a*}\eta^{a}\right]_{3}. \nn
\eea
In (\ref{eq17}), $x^{\mu}$ are the bosonic variables corresponding
to the coordinates of $M$. $g_{\mu \nu}$ are components of the
metric tensor on $M$. $\psi^{\lambda}$ and $\eta^{a}$ are fermionic
real and complex variables associated with the bundles ${\cal S}$
and $V$, respectively. $A_{\mu}^{ab}$ and $F_{\lambda \gamma}^{ab}$
are the components of the connection 1-form and the curvature
2-form on $V$ (written in an orthonormal basis of the structure Lie algebra).
One has the following well-known relation in the Lie algebra:
\be
F_{\mu \nu}=A_{\nu},_{\mu}-A_{\mu},_{\nu} + [A_{\mu},A_{\nu}].
\label{eq19}
\ee
Both $A_{\mu}$ and $F_{\mu \nu}$ in (\ref{eq19}) are antihermitian
matrices. This makes $L$ real up to total time derivatives. In
(\ref{eq17}), $\kappa =0,1$ correspond to switching off and on
of the twisting, respectively. $\alpha$ is a scalar parameter whose
utility will be discussed in Section 4. $\beta$ is the time parameter:
$t\in [0,\beta ]$. Finally, ``dot'' means ordinary time derivative
$\frac{d}{dt}$, and $\frac{D}{dt}$ denotes the covariant time derivative
defined by the Levi Civita connection, e.g.
\[ \frac{D}{dt}\psi^{\gamma}:=\dot{\psi}^{\gamma}+\dot{x}^{\mu}
\Gamma^{\gamma}_{\mu \theta}\psi^{\theta}. \]

The following set of infinitesimal supersymmetric transformations,
\bea
\delta x^{\mu} & = & i \psi^{\mu}\delta \xi  \nonumber \\
\delta \psi^{\gamma} & = &\dot{x}^{\gamma}\delta \xi \label{eq20} \\
\delta \eta^{a} & = & i A_{\gamma}^{ab}\psi^{\gamma}\eta^{b}
\delta \xi \nonumber \\
\delta \eta^{a*} & = & -i A_{\gamma}^{ba}\psi^{\gamma}\eta^{b*}
\delta \xi \nn
\eea
\[ \delta \xi := \mbox{an infinitesimal fermionic variable} \]
leaves the action
\be
S:=\int_{0}^{\beta}L dt
\label{eq21}
\ee
invariant. Hence, the system is supersymmetric.
It is easy to see that the first two equations
in (\ref{eq20}) leave $L_{1}:=[...]_{1}$ in (\ref{eq17}) invariant.
Thus, for $\kappa = \alpha =0$, one has a supersymmetric subsystem.
This subsystem will be  used to compute the index of spin complex
(the Dirac $\hat{A}$ genus) in Section~5. $L_{1}$ can also be
obtained from (\ref{eq1}) by reducing the system of (\ref{eq1}) by
setting $\psi_{1}^{\mu}=\psi_{2}^{\mu}$, for all $\mu$.

The Dynamical equations (\ref{eq12}) are:
\bea
S_{,\t } & = & -g_{\m \t}\frac{D}{dt}\dot{x}^{\m}+i\G_{\l \g \t}\p^{\g}
\frac{D}{dt}\p^{\l}+\mbox{\fs$\frac{i}{2}$}R_{\l \g \s \t}
\dot{x}^{\s}\p^{\g}\p^{\l} + \nn \\
& &  +\k \left[ -iA_{\t}^{ab}(\dot{\e}^{a*}\e^{b}+\e^{a*}\dot{\e}^{b})
+i\dot{x}^{\s}(A_{\s ,\t}^{ab}-A_{\t ,\s}^{ab})\e^{a*}\e^{b}+ \right. \nn \\
& & \left. \hspace{1in} +\mbox{\fs$\frac{1}{2}$}F_{\g \l ,\t}^{ab}
\p^{\g}\p^{\l}\e^{a*}\e^{b} \right] =0 \label{eq22} \\
S_{,\l} & = & g_{\l \g}\frac{D}{dt}\p^{\g}+\k [iF_{\g \l}^{ab}\p^{\g}
\e^{a*}\e^{b}]=0 \nn \\
S_{,a} & = & \k [i\dot{\e}^{a}+i\dot{x}^{\s}A_{\s}^{ab}\e^{b}+
\mbox{\fs$\frac{1}{2}$}F_{\g \l}^{ab}
\p^{\g}\p^{\l}\e^{b}]+\frac{\a}{\b}\e^{a}=0 \nn \\
S_{,b^{*}} & = & \k [-i\dot{\e}^{b*}+i\dot{x}^{\s}A_{\s}^{ab}\e^{a*}+
\mbox{\fs$\frac{1}{2}$}F_{\g \l}^{ab}
\p^{\g}\p^{\l}\e^{a*} ] +\frac{\a}{\b}\e^{b*}=0 \nn
\eea
where the indices from the beginning of the Greek alphabet
($\g ,\d ,\l ,\th ,\e$) label
$\p$'s and those of the middle of the Greek alphabet ($\k ,\cdots$)
label $x$'s, e.g.
\[ S_{,\t} := S\frac{\stackrel{\leftarrow}{\d}}{\d x^{\t}} \hspace{.3in}
\mbox{and}\hspace{.3in}S_{,\l} :=S\frac{\stackrel{\leftarrow}{\d}}{\d
\p^{\l}}. \]
Similarly,
\[ S_{,a} :=S\frac{\stackrel{\leftarrow}{\d}}{\d \e^{a}}\hspace{.3in}
\mbox{and}\hspace{.3in}S_{,a^{*}} := S\frac{\stackrel{\leftarrow}{\d}}{
\d \e^{a*}}. \]

The supersymmetric charge $Q$ corresponding to (\ref{eq20}) is given by
\be
Q \propto g_{\g \m}\p^{\g}\dot{x}^{\m}.
\label{eq23}
\ee
The second functional derivatives of the action are listed in the
following:
\bea
_{\t ,}S_{,\z '} & = & \left[ -g_{\t \z}\frac{\partial^2}{\partial t^2}+
\left( -2\G_{\t \z \m}\dot{x}^{\m}+
\mbox{\fs$\frac{i}{2}$}R_{\l \g \z \t}\p^{\g}\p^{\l}+
\right. \right. \nn  \\
 & & \left. \left. +i\k (A_{\z ,\t}^{ab}-A_{\t ,\z}^{ab})\e^{a*}\e^{b}+
i\G_{\l \g \t}\G^{\l}_{\d \z}\p^{\g}\p^{\d}\right) \pt +C_{\t \z}
\right] \d (t-t') \nn \\
_{\t ,}S_{,\g '} & = &\left[ i\G_{\g \t \d}\p^{\d}\pt +C_{\t \g}\right]
\d (t-t') \nn \\
_{\t ,}S_{,a'} & = &\k \left[ -iA_{\t}^{ca}\e^{c*}\pt +C_{\t a}\right]
\d (t-t') \nn \\
_{\t ,}S_{,a^{*'}} & = &\k \left[ iA_{\t}^{ac}\e^{c}\pt + C_{\t a^{*}}
\right] \d (t-t') \nn \\
_{\l ,}S_{,\z '} & = &\left[ i\G_{\l \z \d}\p^{\d}\pt + C_{\l \z} \right]
\d (t-t') \nn \\
_{\l ,}S_{,\g '} & = &\left[ ig_{\l \g}\pt + C_{\l \g}\right] \d (t-t') \nn \\
_{\l ,}S_{,a'} & = &\k \left[ C_{\l a}\right] \d (t-t') \label{eq24} \\
_{\l ,}S_{,a^{*'}} & = &\k \left[ C_{\l a^{*}}\right] \d (t-t') \nn \\
_{b,}S_{,\z '} & = &\k \left[ -iA_{\z}^{cb}\e^{c*}\pt +C_{b\z}\right]
\d (t-t') \nn \\
_{b,}S_{,\g '} & = &\k \left[ C_{b\g}\right] \d (t-t') \nn \\
_{b,}S_{,a'} & = &\k \left[ C_{ba}\right] \d (t-t') \nn \\
_{b,}S_{,a^{*'}} & = &\k \left[ i\d_{ba}\pt + C_{ba^{*}} \right]
\d (t-t') \nn \\
_{b^{*},}S_{,\z '} & = &\k \left[ iA_{\z}^{bc}\e^{c}\pt + C_{b^{*}\z}
\right] \d (t-t') \nn \\
_{b^{*},}S_{,\g '} & = &\k \left[ C_{b^{*}\g}\right] \d (t-t') \nn \\
_{b^{*},}S_{,a'} & = &\k \left[ i\d_{ba}\pt + C_{b^{*}a}\right]
\d (t-t') \nn \\
_{b^{*},}S_{,a^{*'}} & = &\k \left[ C_{b^{*}a^{*}}\right] \d (t-t')\; . \nn
\eea

In (\ref{eq24}) $C_{ij}$'s \footnote{$\Phi^{i}\in (x^{\m},\p^{\g},
\e^{a},\e^{b*})$ are generic variables.} are terms which do not involve
any time derivative. These terms do not actually contribute to the
equal-time commutation relations of interest, (\ref{eq33}). However, they
will contribute in part to $sdet[G^{+ij}]$ in Section 5. One has:
\bea
C_{\t \z}&:=&-g_{\m \t ,\z}\ddot{x}^{\m}-\G_{\t \m \n ,\z}\dot{x}^{\n}
\dot{x}^{\m} + i\G_{\l \g \t ,\z}\p^{\g}\dot{\p}^{\l} +
i(\G_{\l \g \t}\G^{\l}_{\d \s})_{,\z} \dot{x}^{\s}\p^{\g}\p^{\d}
+ \nn \\
& & +{\mbox \fs \frac{i}{2}}R_{\l \g \s \t ,\z}\dot{x}^{\s}\p^{\g}\p^{\l}
+ \k \ll -iA_{\t ,\z}^{cd}(\dot{\e}^{c*}\e^{d}+\e^{c*}\dot{\e}^{d})+
\right. \nn \\ & & \left.
+i(A_{\s,\t \z}^{cd}-A_{\t ,\s \z}^{cd})\dot{x}^{\s}\e^{c*}\e^{d} +
\mbox{\fs$\frac{1}{2}$}F_{\g \l ,\t \z}^{cd}\p^{\g}\p^{\l}
\e^{c*}\e^{d} \rr \nn \\
C_{\t \g}&:=&-i\G_{\l \g \t}\dot{\p}^{\l}+i(\G_{\l \d \t}\G^{\l}_{\g \s}
-\G_{\l \g \t}\G^{\l}_{\d \s})\dot{x}^{\s}\p^{\d}+iR_{\g \l \d \t}
\dot{x}^{\s}\p^{\l} + \nn \\
& &  +\k \ll F_{\l \g ,\t}^{cd}\p^{\l}\e^{c*}\e^{d} \rr \nn \\
C_{\t a}&:=&-iA_{\t}^{ca}\dot{\e}^{c*}+i(A_{\s ,\t}^{ca}-A_{\t ,\s}^{ca})
\dot{x}^{\s}\e^{c*}+ \mbox{\fs$\frac{1}{2}$}F_{\g \l ,\t}^{ca}\p^{\g}\p^{\l}
\e^{c*} \nn \\
C_{\t a^{*}}&:=&iA_{\t}^{ac}\dot{\e}^{c}-i(A_{\s ,\t}^{ac}-
A_{\t ,\s}^{ac})\dot{x}^{\s}\e^{c} + \mbox{\fs$\frac{1}{2}$}
F_{\g \l}^{a^{*}c}\p^{\g}\p^{\l}\e^{c} \nn \\
C_{\l \z}&:=&i\G_{\l \d \m ,\z}\dot{x}^{\m}\p^{\d}+ig_{\l \d ,\z}
\dot{\p}^{\d}-\k \ll F_{\d \l ,\z}^{cd}\p^{\d}\e^{c*}\e^{d} \rr \nn \\
C_{\l \g}&:=&i\G_{\l \g \m}\dot{x}^{\m}+\k F_{\l \g}^{ab}\e^{a*}\e^{b}
\nn \\
C_{\l a}&:=&-F_{\d \l}^{ca}\p^{\d}\e^{c*} \label{eq25} \\
C_{\l a^{*}}&:=&F_{\d \l}^{ac}\p^{\d}\e^{c} \nn \\
C_{b\z}&:=&-iA_{\s ,\z}^{cb}\dot{x}^{\s}\e^{c*}-\mbox{\fs$\frac{1}{2}$}
F_{\g \l}^{cb}\p^{\g}\p^{\l}\e^{c*} \nn \\
C_{b\g}&:=&F_{\d \g}^{cb}\p^{\d}\e^{c*} \nn \\
C_{ba}&:=&0 \nn \\
C_{ba^{*}}&:=&-iA_{\s}^{ab}\dot{x}^{\s}-\mbox{\fs$\frac{1}{2}$}
F_{\d \th}^{ab}\p^{\d}\p^{\th}-\mbox{\fs$\frac{\a}{\k \b}$}\d^{ab} \nn \\
C_{b^{*}\z}&:=&iA_{\s ,\z}^{bc}\dot{x}^{\s}\e^{c}+
\mbox{\fs$\frac{1}{2}$}F_{\g \l ,\z}^{bc}\p^{\g}\p^{\l}\e^{c} \nn \\
C_{b^{*}\g}&:=&-F_{\d \g}^{bc}\p^{\d}\e^{c} \nn \\
C_{b^{*}a}&:=&iA_{\s}^{ba}\dot{x}^{\s}+\mbox{\fs$\frac{1}{2}$}
F_{\d \th}^{ba}\p^{\d}\p^{\th}+\mbox{\fs$\frac{\a}{\k \b}$}\d^{ba} \nn \\
C_{b^{*}a^{*}}&:=&0 .\nn
\eea

The advanced Green's functions $G^{+ij'}$ are calculated using
(\ref{eq13}) and (\ref{eq24}). The results are listed below:
\renewcommand{\tt}{\theta (t'-t)}
\bea
G^{+\xi \z '} & = &\tt \ll -g^{\xi '\z'}(t-t')+
\mbox{\fs$\frac{1}{2}$}g^{\xi '\t '}
\left( 2\G_{\t '\n '\m '}\dot{x}^{\m '}- \mbox{\fs$\frac{i}{2}$}
R_{\n '\t '\l '\g '}\p^{\g '}\p^{\l '}-  \right. \right.  \nn \\
& &  \left. \left. +i\k F_{\t '\n '}^{a'b'}\e^{a*'}\e^{b'} \right) g^{\n '\z '}
(t-t')^{2}+O(t-t')^{3} \rr \nn \\
G^{+\xi \g '} & = &\tt \ll g^{\xi '\t '}\G^{\g '}_{\t '\l '}\p^{\l '}
(t-t')+O(t-t')^{2} \rr \nn \\
G^{+\xi a'} & = &\tt \ll g^{\xi '\t '}A_{\t '}^{a'c'}\e^{c'}(t-t')+
O(t-t')^{2}\rr \nn \\
G^{+\xi a^{*'}} & = &\tt \ll g^{\xi '\t '}A_{\t'}^{c'a^{*'}}\e^{c*'}
(t-t')+O(t-t')^{2} \rr \nn \\
G^{+\l \z '} & = &\tt \ll g^{\z '\t '}\G^{\l '}_{\t '\d '}\p^{\d '}
(t-t')+O(t-t')^{2} \rr \nn \\
G^{+\l \g '} & = &\tt \ll -ig^{\l '\g'}+O(t-t') \rr \nn \\
G^{+\l a'} & = &\tt \ll O(t-t') \rr \label{eq26} \\
G^{+\l a^{*'}} & = &\tt \ll O(t-t') \rr \nn \\
G^{+b\z '} & = &\tt \ll g^{\z '\t '}A_{\t '}^{b'c'}\e^{c'}(t-t')+O(t-t')\rr
\nn \\
G^{+b\g '} & = &\tt \ll O(t-t')\rr \nn \\
G^{+ba'} & = &\tt \ll O(t-t') \rr \nn \\
G^{+ba^{*'}} & = &\tt \ll - \mbox{\fs$\frac{i}{\k}$}\d^{b'a'}+
O(t-t')\rr \nn \\
G^{+b^{*}\z '} & = &\tt \ll -g^{\z '\t '}A_{\t '}^{c'b'}\e^{c*'}
(t-t')+O(t-t')^{2}\rr \nn \\
G^{+b^{*}\g '} & = &\tt \ll O(t-t')\rr  \nn \\
G^{+b^{*}a'} & = &\tt \ll - \mbox{\fs$\frac{i}{\k}$}\d^{b'a'}+O(t-t')\rr \nn \\
\nopagebreak[3]  G^{+b^{*}a^{*'}} & = &\tt \ll O(t-t')\rr \; .\nn
\eea

The Green's functions $G^{-ij'}$ and $\tilde{G}^{ij'}$ are then obtained
using (\ref{eq15.1}) and (\ref{eq14}), respectively. Substituting the latter
in (\ref{eq15}) leads to the following Peierls brackets:
\bea
(x^{\xi},x^{\z '}) & = &-g^{\xi '\z '}(t-t')+
\mbox{\fs$\frac{1}{2}$}g^{\xi '\t'} \left(
2\G_{\t '\n '\m '}\dot{x}^{\m '}-
\mbox{\fs$\frac{i}{2}$}R_{\n '\t '\l '\g '}
\p^{\g '}\p^{\l '}+ \right. \nn \\
& & \left. -i\k F_{\t '\n '}^{a'b'}\e^{a^{*'}}\e^{b'} \right)
g^{\n '\z '}(t-t')^{2}+O(t-t')^{3} \nn \\
(x^{\xi},\p^{\g '}) & = &g^{\xi '\t '}\G^{\g '}_{\t '\d '}\p^{\d '}
(t-t')+O(t-t')^{2} \nn \\
(x^{\xi},\e^{a'}) & = &g^{\xi '\t '}A_{\t '}^{a'c'}\e^{c'}(t-t')+
O(t-t')^{2} \label{eq28} \\
(x^{\xi},\e^{a*'}) & = &-g^{\xi '\t '}A_{\t '}^{c'a'}\e^{c*'}(t-t')
+O(t-t')^{2} \nn \\
(\p^{\l},\p^{\g '}) & = &-ig^{\l '\g '}+O(t-t') \nn \\
(\e^{b},\e^{a*'}) & = &-\mbox{\fs$\frac{i}{\k}$}\d^{b'a'}+O(t-t')\; . \nn
\eea
Other possible Peierls brackets are all of the order $(t-t')$ or
higher. Differentiating the necessary expressions in (\ref{eq28})
with respect to $t$ and $t'$, one arrives at the Peierls brackets
among the coordinates $(x,\p ,\e ,\e^{*})$ and their time derivatives.
The interesting equal-time Peierls brackets are the following:
\bea
(x^{\xi},x^{\z}) &=&(x^{\xi},\p^{\g}) = (x^{\xi},\e^{a}) =
(x^{\xi},\e^{a*}) = 0 \nn \\
(\p^{\g},\e^{a}) &=&(\p^{\g},\e^{a*}) = (\e^{a},\e^{b}) = (\e^{a*},\e^{b*})
 = 0 \nn \\
(\p^{\l},\p^{\g}) &=&-ig^{\l \g} \label{eq29} \\
(\e^{a},\e^{b*}) &=&-\mbox{\fs$\frac{i}{\k}$}\d^{ab} \nn \\
(\dot{x}^{\xi},x^{\z}) &=&-g^{\xi \z} \nn \\
(\dot{x}^{\xi},\dot{x}^{\z}) &=&g^{\xi \t}\left[ (\G_{\n \t \m}-
\G_{\t \n \m})\dot{x}^{\m}+
\mbox{\fs$\frac{i}{2}$}R_{\t \n \l \g}\p^{\l}\p^{\g}
+i\k F_{\t \n}^{ab}\e^{a*}\e^{b}\right] g^{\n \z} \nn \\
(\dot{x}^{\xi},\p^{\g}) &=&g^{\xi \t}\G^{\g}_{\t \l}\p^{\l} \nn \\
(\dot{x}^{\xi},\e^{a}) &=&g^{\xi \t}A_{\t}^{ac}\e^{c} \nn \\ \nopagebreak[3]
(\dot{x}^{\xi},\e^{a*}) &=&-g^{\xi \t}A_{\t}^{ca}\e^{c*} \; .\nn
\eea
The next step is to define the appropriate momenta conjugate to $x^{\n}$.
The canonical momenta are given by:
\be
p_{\n}^{\rm (canonical)} :=L\frac{\stackrel{\leftarrow}{\partial}}{\partial
\dot{x}^{\n}}=g_{\n \m}\dot{x}^{\m}+
\mbox{\fs$\frac{i}{2}$}\G_{\l \g \n}\p^{\l}\p^{\g}+
i\k A_{\n}^{ab}\e^{a*}\e^{b}.
\label{eq30}
\ee
A more practical choice is provided by:
\be
p_{\n}:=g_{\n \m}\dot{x}^{\m} \; .
\label{eq31}
\ee
This choice together with the use of (\ref{eq14.9}) and (\ref{eq29})
lead to:
\bea
(p_{\n},x^{\m}) & = &-\d_{\n}^{\m}  \nn \\
(p_{\n},\p^{\g}) & = &\G^{\g}_{\n \l}\p^{\l}  \nn  \\
(p_{\n},\e^{a}) & = &A_{\n}^{ac}\e^{c} \label{eq32} \\
(p_{\n},\e^{a*}) & = &-A_{\n}^{ca}\e^{c*} \nn \\
(p_{\n},p_{\m}) & = &\mbox{\fs$\frac{i}{2}$}R_{\n \m \l \g}\p^{\l}\p^{\g}
+\k \ll iF_{\n \m}^{ab}\e^{a*}\e^{b} \rr.  \nn
\eea
The quantization is performed via (\ref{eq16}). Enforcing (\ref{eq16})
and using (\ref{eq29}) and (\ref{eq32}), one has the following
supercommutation relations. For convenience, the commutators,
$[ . , .]$, and the anticommutators, $\{ . , .\}$, are distinguished.
\bea
[x^{\m},x^{\n}] &=&[x^{\m},\p^{\g}] = [x^{\m},\e^{\a}] = [x^{\m},\e^{a*}]
= 0 \nn \\
\{\p^{\l},\e^{a}\} &=&\{\p^{\l},\e^{a*}\} = \{\e^{a},\e^{b}\} =
\{\e^{a*},\e^{b*}\} = 0 \nn \\
\{\p^{\l},\p^{\g}\} &=& \hbar g^{\l \g} \nn \\
\{\e^{a},\e^{b*}\} &=& \frac{\hbar}{\k}\d^{ab^{*}} \nn \\
\ll x^{\m},p_{\n}\rr &=& i\hbar \d^{\m}_{\n} \label{eq33} \\
\ll \p^{\g},p_{\n} \rr &=& -i\hbar \G^{\g}_{\n \l}\p^{\l} \nn \\
\ll \e^{a},p_{\n} \rr &=& -i\hbar A_{\n}^{ac}\e^{c} \nn \\
\ll \e^{a*},p_{\n} \rr &=& i\hbar A_{\n}^{ca}\e^{c*} \nn \\
\ll p_{\m},p_{\n} \rr &=& - \mbox{\fs$\frac{\hbar}{2}$}R_{\m \n \l \g}
\p^{\l}\p^{\g}  +\k \ll - \mbox{\fs$\hbar$}F_{\m \n}^{ab}\e^{a*}\e^{b} \rr .
\nn
\eea
One must note that in general there may be  factor ordering ambiguities in
the right hand side of (\ref{eq16}). Indeed, for the example considered
in this paper there are  three inequivalent
choices for the last equation in (\ref{eq33}). These correspond to
the following choices of ordering $\e^{a*}\e^{b}$ in (\ref{eq32}):
\be
\e^{a*}\e^{b} \hspace{.15in},\hspace{.15in} -\e^{b}\e^{a*}
\hspace{.15in},\:{\rm and} \hspace{.15in} \mbox{\fs$\frac{1}{2}$}
(\e^{a*}\e^{b}-\e^{b}\e^{a*}).
\label{eq33.1}
\ee
The first choice is selected in
(\ref{eq33}) because, as will be seen in Section 4, it leads to the
identification of the supersymmetric charge with the twisted
Dirac operator.
The quantum mechanical supersymmetric charge corresponding to (\ref{eq23}),
which is also hermitian, is given by:
\be
Q=\mbox{\fs$\frac{1}{\sqrt{\hbar}}$}
\p^{\n}g^{\frac{1}{4}} p_{\n} g^{-\frac{1}{4}}.
\label{eq34}
\ee
Hermiticity is ensured in view of the identity:
 \[ \p^{\n}g^{\frac{1}{4}}p_{\n}g^{-\frac{1}{4}}=g^{-\frac{1}{4}}
p_{\n}g^{\frac{1}{4}}\p^{\n}. \]
Here, $g$ is the determinant of $(g_{\m \n})$ and the proportionality
constant, $1/\sqrt{\hbar}$, is fixed by comparing the reduced form
of (\ref{eq35}) to the case: $\p =\e =0$ and $M=\R^{n}$. Equation
(\ref{eq34}) together with (\ref{eq10.1}) yield the Hamiltonian:
\be
H = Q^{2} =
\mbox{\fs$\frac{1}{2}$}g^{-\frac{1}{4}}p_{\m}g^{\frac{1}{2}}g^{\m \n}
p_{\n}g^{-\frac{1}{4}} + \mbox{\fs$\frac{\hbar^{2}}{8}$}R +\k \ll
-\mbox{\fs$\frac{1}{2}$}F_{\l \g}^{ab}\p^{\l}\p^{\g}\e^{a*}\e^{b} \rr \; .
\label{eq35}
\ee
The derivation of (\ref{eq35}) involves repeated use of (\ref{eq33}). In
particular, the appearence of
the scalar curvature term, $\frac{\hbar^{2}}{8}R$, is a
consequence of the third and the last equations in (\ref{eq33})
and the symmetries of the Riemann curvature tensor. Reducing the system of
(\ref{eq17}) to a purely bosonic one, i.e. setting $\p =\e =0$, leads
to the problem of the dynamics of a free particle moving on a Riemannian
manifold. Equation (\ref{eq35}) is in complete agreement with the analysis of
the latter problem by Bryce DeWitt \cite{bd1}. One has to emphasize,
however, that here the Hamiltonian is obtained as a result of the
superalgebra condition (\ref{eq10.1}). One can also ckeck that
(\ref{eq35}) reduces to the classical Hamiltonian, i.e. $L\frac{
\stackrel{\leftarrow}{\partial}}{\partial \dot{\Phi}^{i}}\dot{\Phi}^{i}-L$,
as $\hbar \rightarrow 0$, for the Lagrangian (\ref{eq17}) with
$\alpha = 0$. It turns out that this is precisely what one needs
for the proof of the twisted spin index theorem. See Section 4, for a more
detailed discussion of this point.

\section{The Quantum System}
In the rest of this paper $\hbar$ will be set to 1.
\subsection{The Case of Spin Complex ($\k =\a = 0$)}

Let $\{e^{i}_{\m}dx^{\m} \}$ be a local orthonormal frame for the
cotangent bundle, $TM^{*}$, i.e. $e^{i}_{\m}e^{j}_{\n}\d_{ij}=g_{\m
\n}$ and $\{e^{\m}_{i}\frac{\partial}{\partial x^{\m}}\}$ be its
dual in $TM$. Consider,
\be
\g^{i}:=i\sqrt{2} e^{i}_{\m}\p^{\m}.
\label{eq36}
\ee
Then, (\ref{eq33}) implies:
\be
\{ \g^{i},\g^{j}\}=-2\d_{ij}
\label{eq37}
\ee
In the mathematical language one says that $\g^{i}$'s are the
generators of the Clifford algebra ${\cal C}(TM_{x}^{*})\otimes \C$,
\cite[Part I\hspace{-.4mm}I, p. 6]{cd1}. Furthermore, for $i=1,...,l:=m/2$
define \cite{windey,manes-zumino}:
\bea
\xi^{i}& := & \frac{1}{2}(\g^{2i-1}+i\g^{2i}) \nn \\
\xi^{i\dagger}& := & -\frac{1}{2}(\g^{2i-1}-i\g^{2i}).
\label{eq38}
\eea
Equations (\ref{eq37}) and (\ref{eq38}) yield the following anticommutation
relations:
\bea
\{\xi^{i},\xi^{j\dagger}\}&=&\d^{ij} \nn \\
\{\xi^{i},\xi^{j}\}&=&\{\xi^{i\dagger},\xi^{j\dagger}\} = 0.
\label{eq39}
\eea
(\ref{eq39}) indicates that $\xi^{i}$ and $\xi^{i\dagger}$ behave as
fermionic annihilation and creation operators. The basic kets of the
corresponding Fock space are of the form:
\be
\mid i_{r},...,i_{1},x,t \kt :=\xi^{i_{r}\dagger}...\xi^{i_{1}\dagger}
\mid\! x,t \kt \; .
\label{eq40}
\ee
The wavefunctions are given by:
\be
\Psi_{i_{1},...,i_{r}}(x,t)=\br x,t,i_{1},...,i_{r}\mid\! \Psi \kt
\label{eq41}
\ee
where
\[ \br x,t,i_{1},...,i_{r}\mid := \mid\! i_{r},...,i_{1},x,t
\kt^{\dagger}. \]

The chirality operator $(-1)^{f}$ of (\ref{eq11}) is defined by
\be
\g^{m+1}:=i^{l}\g^{1}...\g^{m}=\prod_{i=1}^{l}(1-2\xi^{i\dagger}
\xi^{i})\; .
\label{eq42}
\ee
Equation (\ref{eq42}) is a clear indication of the relevance of the
system to the spin complex. In fact, in terms of $\g$'s the
supersymmetric charge, (\ref{eq34}), is written as
\be
Q=\mbox{\fs$\frac{-i}{\sqrt{2}}$}g^{\frac{1}{4}}\g^{\n}p_{\n}g^{-\frac{1}{4}},
\label{eq43}
\ee
where
\be
\g^{\n}:=e^{\n}_{i}\g^{i}.
\label{eq44}
\ee
It is not difficult to see that indeed $Q$ is represented by the Dirac
operator $\not\!\! D$ in the coordinate representation, i.e.
\be
\br x,t,i_{1},...,i_{r}\mid p_{\m}=-i\not\!\partial_{\m}
\br x,t,i_{1},...,i_{r}\mid
\label{eq45}
\ee
with
\be
\not\!\partial_{\m}:=\frac{\partial}{\partial x^{\m}}-
\mbox{\fs$\frac{1}{8}$}\omega_{\m}.
\label{eq45.1}
\ee
The following commutation relations can be easily computed:
\bea
\ll x^{\m},-i\not\!\partial_{\n}\rr & = & i\d^{\m}_{\n} \label{eq46} \\
\ll \g^{i},-i\not\!\partial_{\n}\rr & = & -i\omega^{i}_{j\n}\g^{j}
\label{eq47} \\
\ll-i\not\!\partial_{\m},-i\not\!\partial_{\n}\rr & = &
\mbox{\fs$\frac{1}{4}$}R_{\m \n \l \d}\g^{\l}\g^{\d} \; .\label{eq48}
\eea
In (\ref{eq45.1}) and (\ref{eq47}) $\omega_{\m}$ and $\omega^{i}_{j\n}$
refer to the spin connection:
\be
\omega^{i}_{j\n}:=\G^{\m}_{\n \s}e^{i}_{\m}e^{\s}_{j}-e^{i}_{\m ,\n}
e^{\m}_{j}=:\omega_{\n ij}\; .
\label{eq49}
\ee
\be
\omega_{\m}:=\omega_{\m ij}[\g^{i},\g^{j}]=2\omega_{\m ij}\g^{i}\g^{j}
\label{eq50}
\ee
In the derivation of (\ref{eq48}) one uses the symmetries of
$\omega_{\n ij}$, especially the identity:
\[ \omega_{\n ij}=-\omega_{\n ji}. \]
Comparing (\ref{eq46}), (\ref{eq47}), (\ref{eq48}) with the last three
equations in (\ref{eq33}) justifies (\ref{eq45}) and the claim
preceding it.

Following the analysis of \cite[\S 6.7]{bd1}, the coherent state
representation can be used to give a path integral representation
of the supertrace of any operator, $\hat{O}$. The following relations
summarize this procedure. The coherent states are defined by
\bea
\mid\! x,\xi ;t\kt &:=&e^{-\frac{1}{2}\xi^{j*}\xi^{j}+\hat{\xi}^{j\dagger}(t)
\xi^{j}}\mid\! x,t\kt \nn \\
\br x,\xi^{*};t\mid & := & \mid\! x,\xi ;t\kt^{\dagger}
\hspace{.3in} , \hspace{.3in} \xi \in \C \; .
\label{eq51}
\eea
The `` $\hat{ }$ '' is used to distinguish the operators from the
scalars where necessary. Equation (\ref{eq51}) leads to
\be
str(\hat{O})=\frac{1}{(2\pi i)^{l}}\int \br x,\xi^{*};t\mid \hat{O}
\mid\! x,\xi ;t \kt d^{m}\!\!x\: d^{l}\xi^{*}\: d^{l}\xi .
\label{eq53}
\ee
In particular, one has
\be
str(e^{-i\b H})=\frac{1}{(2\pi i)^{l}}\int \br x,\xi^{*};t+\b \mid\!
x,\xi ;t \kt d^{m}\!\!x\: d^{l}\xi^{*}\: d^{l}\xi.
\label{eq54}
\ee
The following notation is occasionally used:
\be
K(x,\xi ;\b ) :=\br x, \xi^{*} ;\b \mid x,\xi ;0 \kt
\label{eq55}
\ee
(\ref{eq55}) has a well-known path integral representation. One can change
the variables $\xi$'s to $\p$'s in (\ref{eq54}) and (\ref{eq55}) to
compute the index. This will be pursued in Section 6.

\subsection{The Case of the Twisted Spin Complex ($\k =1$)}
The commutation relations between $\e$'s and $\e^{*}$'s in (\ref{eq33}),
with $\hbar =1$ and $\e^{\dagger}:=\e^{*}$, read
\bea
\{ \e^{a},\e^{b\dagger} \}&=&\d^{ab} \nn \\
\{ \e^{a},\e^{b} \}&=&\{ \e^{a\dagger},\e^{b\dagger} \} = 0 \; .
\label{eq56}
\eea
Thus, $\e$ and $\e^{\dagger}$ can be viewed as the annihilation and
creation operators for ``$\e$-fermions''. The total Fock space ${\cal
F}_{\rm tot.}$ is the tensor product of the Fock space ${\cal F}_{0}$
of the $\k =0$ case and the one constructed by the action of
$\e^{\dagger}$'s on the vacuum. The basic kets are:
\[ \mid\! a_{p},...,a_{1},i_{r},...,i_{1},x,t\kt :=
\mid\! a_{p},...,a_{1},x,t\kt \otimes \mid\! i_{r},...,i_{1},x,t\kt , \]
where
\[ \mid\! a_{p},...,a_{1},x,t\kt :=\e^{a_{p}\dagger}...\e^{a_{1}\dagger}
\mid\! x,t\kt . \]
The relevant Fock space for the twisted spin complex, however, is the
subspace ${\cal F}_{V}$ of ${\cal F}_{\rm tot.}$ spanned by the
1-$\e$-particle state vectors. These are represented by the following
basic kets:
\be
\mid a,i_{r},...,i_{1},x,t\kt .
\label{eq57}
\ee
In the coordinate representation one has:
\be
\br x,t,i_{1},...,i_{r},a \mid p_{\m} =
-i(\not\!\partial_{\m}+{\cal A}_{\m})
\br x,t,i_{1},...,i_{r},a \mid ,
\label{eq56.1}
\ee
where
\[{\cal A}_{\m} := A_{\m}^{ab}\e^{a*}\e^{b} .\]
This is justified by computing
the following commutation relations and comparing them with the
last five relations in (\ref{eq33}):
\bea
\ll x^{\m},-i(\not\!\partial_{\n}+{\cal A}_{\n}) \rr &=&i\d^{\m}_{\n} \nn \\
\ll \g^{i},-i(\not\!\partial_{\n}+{\cal A}_{\n}) \rr &=&
-i\omega^{i}_{j\n}\g^{j} \nn \\
\ll \e^{a},-i(\not\!\partial_{\n}+{\cal A}_{\n}) \rr &=&
-iA_{\n}^{ab}\e^{b} \nn \\
\ll \e^{a*},-i(\not\!\partial_{\n}+{\cal A}_{\n}) \rr &=&
iA_{\n}^{ca}\e^{c*} \nn \\
\ll -i(\not\!\partial_{\m}+{\cal A}_{\m}) ,
-i(\not\!\partial_{\n}+{\cal A}_{\n}) \rr &=& \mbox{\fs$\frac{1}{4}$}
R_{\m \n \l \d}\g^{\l}\g^{\d}-F_{\m \n}^{ab}\e^{a*}\e^{b} \; . \nn
\eea
Again, the supersymmetric charge $Q$ of (\ref{eq34}) is identified with
the twisted Dirac operator, $\not\!\! D_{V}$, in the coordinate representation.
In view of (\ref{eq34}), (\ref{eq36}), and (\ref{eq56.1}), one has
\be
\br x,t,i_{1},...,i_{r},a\!\mid Q =
\mbox{\fs$\frac{-1}{\sqrt{2}}$}
g^{\frac{1}{4}}\ll \g^{\m}(\not\!\partial_{\m}+{\cal A}_{\m})\rr
g^{-\frac{1}{4}} \br x,t,i_{1},...,i_{r},a\!\mid .
\label{eq58}
\ee

Once more, $\g^{m+1}$ of (\ref{eq42}) serves as the chirality operator.
In particular, $Q$ switches the $\pm 1$-eigenspaces of $\g^{m+1}$, i.e.
$\{ \g^{m+1},Q\} = 0$.

Coherent states are defined by
\[ \mid\! x,\xi ,\e ;t\kt :=e^{-\frac{1}{2}\e^{a*}\e^{a}+
\hat{\e}^{a\dagger}\e^{a}}\mid\! x,\xi ;t\kt . \]
The supertrace formula, the analog of (\ref{eq53}), is given by:
\[ str(\hat{O})=\frac{1}{(2\pi i)^{l+n}}\int \br x,\xi^{*},\e^{*};t\mid
\hat{O} \mid\! x,\xi ,\e ;t\kt\, d^{m}\!x\, d^{l}\!\xi^{*} d^{l}\!\xi
\, d^{n}\!\e^{*} d^{n}\!\e .\]
The application of the latter equation to the time evolution operator
leads to
\be
str(e^{-i\b H})=\frac{1}{(2\pi i)^{l+n}}\int \br x,\xi^{*},\e^{*};t+\b
\mid\! x,\xi ,\e ;t\kt d^{m}\!x\, d^{l}\!\xi^{*} d^{l}\!\xi\, d^{n}\!\e^{*}
d^{n}\!\e .
\label{eq59}
\ee
Equation (\ref{eq59}) does not, however, provide the index. This is
because in (\ref{eq59}) the supertrace is taken over
${\cal F}_{\rm  tot.}$, rather than ${\cal F}_{V}$. This is remedied
by including a term of the form
\newcommand{\eno}{e^{i\a \hat{\e}^{a\dagger}\hat{\e}^{a}}}
\[ \eno , \]
in (\ref{eq59}), and considering
\be
str \ll e^{-i\b H} \eno \rr .
\label{eq61}
\ee
The linear term in
\[ \l := e^{i\a} \]
in (\ref{eq61}) is precisely the index of
$\not\!\partial_{V}$.\footnote{Note that (\ref{eq61}) is a polynomial
in $\l$} The term $[...]_{3}$ in the original
Lagrangian (\ref{eq17}) is  added to fulfill this objective,
\cite{windey,manes-zumino}. In Section 7,
\be
H_{\rm eff.}:= H - \mbox{\fs$\frac{\a}{\b}$}
\hat{\e}^{a\dagger}\hat{\e}^{a}
\label{eq61.1}
\ee
will be used in the path integral evaluation of the kernel:
\be
K(x,\xi ,\e ;\b ):=\br x,\xi^{*},\e^{*};\b \mid\! x,\xi ,\e ;0\kt .
\label{eq62}
\ee
The index of $\not\!\partial_{V}$ is then given by
\be
index(\not\!\partial_{V})=\left. \frac{\partial}{\partial \l}\right|_{\l =0}
 str(e^{-i\b H_{\rm eff.}}) \hspace{.1in} .
\label{eq63}
\ee

\section{The Path Integral Evaluation of the Kernel, the Loop Expansion
and the Green's Function Methods}

The path integral evaluation of the kernel, (\ref{eq64}) below, is
discussed in \cite[\S 5]{bd1}. In general, for a quadratic Lagrangian the
following relation holds:
\be
K(\Phi '',t''\mid \Phi ',t'):=\br \Phi '',t''\mid \Phi ',t'\kt
=Z\int_{(\Phi ',t')}^{(\Phi '',t'')} e^{iS[\Phi ]}\left( sdet
G^{+}[\Phi ]\right)^{-\frac{1}{2}}{\cal D}\Phi .
\label{eq64}
\ee
Here , $Z$ is a (possibly infinite) constant of functional integration.
In the loop expansion of (\ref{eq64}), one expands the field
(coordinate) variables around the (classical) solutions of the dynamical
equations, $\Phi_{0}(t)$:
\be
\Phi^{i}(t)=\Phi^{i}_{0}(t)+\f^{i}(t).
\label{eq65}
\ee
Substituting (\ref{eq65}) in (\ref{eq64}) and expanding in power series
around $\Phi_{0}$, one has~\footnote{ There are additional
complications if $M$ has nontrivial first homology group, \cite{bd1}.
However, this is not relevant to the computation of the index. See
Section 8 for a further discussion of this problem. }~:
\be
K(\Phi '',t''\mid \Phi ',t')=Z (sdet G_{0}^{+})^{-\frac{1}{2}}
e^{i\so }\int e^{\frac{i}{2}\f^{i}\: _{i,}\! S_{0 ,j} \f^{j}} \{ 1+\cdots \}
{\cal D}\f \; .
\label{eq66}
\ee
Here,``$\cdots$'' denotes the higher order terms starting with the 2-loop
terms \footnote{The 2-loop terms will be analyzed in \cite{R-factor}.
They are of order $\b$ or higher. }.
The subscript ``$_{0}$'' means that the corresponding quantity is
evaluated at the classical solution $\Phi_{0}$, e.g.
\[ _{i,}S_{0 ,j}:=\: _{i,}\! S_{,j}[\Phi_{0}] . \]
The lowest order approximation
of (\ref{eq64}), which is explicitly shown in (\ref{eq66}), is the
well-known WKB approximation. This term can be further simplified
if the surviving Gaussian functional integral in (\ref{eq66}) is
evaluated. Generalizing the ordinary Gaussian integral formula,
one has \footnote{In (\ref{eq67}) $c$ is a constant of functional integration,
it may be identified with 1 if the action is appropriately rescaled.
However, this does not play any role in the application of (\ref{eq67})
and thus is not pursued here.}:
\be
\int e^{\frac{i}{2}\f^{i}\: _{i,}\! S_{0 ,j}\f^{j}} {\cal D}\f =
c [sdet(G)]^{\frac{1}{2}} \; .
\label{eq67}
\ee
The Green's function $G=(G^{ij})$, which appears in (\ref{eq67}), is the
celebrated Feynman propagator. It is the inverse of $(  _{j,}S_{0 ,i} )$, :
\be
_{i,}S_{0 ,j'}G^{j'k''} = -\d_{j}^{k}\d (t-t'') ,
\label{eq68}
\ee
defined by the boundary conditions which fix the end points:
\be
\Phi (t')=\Phi ' \hspace{.2in}\mbox{and}\hspace{.2in} \Phi (t'')=\Phi '' .
\label{eq68.1}
\ee
An important property of $G^{ij}$ is the following \cite{bd1}:
\be
G^{ij'}=(-1)^{ij'}G^{j'i} .
\label{eq68.2}
\ee
The emergence of $sdet(G^{+})$ in (\ref{eq64}) is quite important and
must not be underestimated. An explicit computation of $sdet(G^{+})$
for (\ref{eq17}) is in order. The main tool is the definition of the
$sdet$ as the solution of the variational equation \cite[\S 1]{bd1}:
\be
\d \ln [sdet(G^{+})] := str[(G^{+})^{-1} \d G^{+}] \; .
\label{eq69}
\ee
Using the definition of $G^{+}$ (\ref{eq13}):
\be
(G^{+ij})^{-1} := -( _{i,}S_{,j})
\label{eq70}
\ee
one has
\newcommand{\gr}{G^{+}\left|_{\Phi ',t'}^{\Phi '',t''}\right.}
\newcommand{\ddt}{\d \ln\ll sdet\left( \gr \right)\rr}
\be
\ddt =\int_{t'}^{t''}d\t\int_{t'}^{t''}d\t'
(-1)^{i}\: _{i,}\d S_{,j'}G^{+j'i} .
\label{eq71}
\ee
In (\ref{eq71}), the variation in the action is with respect to
the functional variation of the metric and the connection fields, i.e.
$\d g_{\m \n}$ and $\d A_{\m}^{ab}$, respectively.
In view of (\ref{eq24}) and (\ref{eq26}), equation (\ref{eq71}) becomes:
\newcommand{\dt}{sdet\left( G^{+}\mid_{\Phi',t_{0}'}^{\Phi'',t_{0}''}\right)}
\bea
\d \ln \dt &=& \int_{t'}^{t''}dt\ll \frac{d}{dt}(\d \ln g)
-iG^{+\g \l}_{1}\d g_{\l \g}+ \right. \label{eq72} \\
& & \hspace{.5cm} \left. i\d^{ab}(\d C_{ba^{*}}+ \d C_{b^{*}a})+
ig^{\g \l}\d C_{\l \g}\rr \th (0) .\nn
\eea
In (\ref{eq72}), $G^{+\g \l}_{1}$ are the coefficients of the linear terms
in the expansion of $G^{+\g \l}$ in (\ref{eq26}), i.e.
\be
G^{+\g \l '} =: \th (t-t')\ll -ig^{\g '\l '}+G^{+\g '\l '}_{1}
(t-t')+O(t-t')^{2}\rr ,
\label{eq73}
\ee
and $\d C$'s are the variations of the corresponding terms in (\ref{eq25}).
It is quite remarkable that although other higher order terms in
(\ref{eq26}) originally enter in (\ref{eq71}), their contributions
cancel and one is finally left with (\ref{eq72}). The fortunate
cancellations seem to be  primarily due to supersymmetry. Incidentally, the
calculation of $G^{+\g \l}_{1}$ is quite straightforward. The 16 coupled
equations, (\ref{eq13}), which give the next order terms in (\ref{eq26}),
decouple miraculously to yield:
\be
G^{+\g \l}_{1}=g^{\d \g}g^{\th \l}C_{\d \th}-g^{\z \t}\G^{\g}_{\z \d}
\G^{\l}_{\t \th}\p^{\d}\p^{\th} .
\label{eq74}
\ee
Substitituing (\ref{eq74}) in (\ref{eq72}), the second term in (\ref{eq74})
drops after contracting with $\d g_{\l \g}$. The surviving term combines
with the last term in (\ref{eq72}) to produce a factor of
\[ i\d (g^{\g \l}C_{\l \g}) . \]
Finally, using (\ref{eq25}) and the identity
\[ \G^{\g}_{\g \m}=\mbox{\fs$\frac{1}{2}$}\partial_{\m}(\ln g) \]
one obtains:
\be
\d \ln \dt =\int_{t'}^{t''}\d \ll \mbox{\fs$\frac{1}{2}$}
(\frac{d}{dt}\ln g) \rr \th (0) dt .
\label{eq75}
\ee
Adapting $\th (0)=1/2$, \cite[\S 6.4]{bd1}, and integrating (\ref{eq75}),
one has:
\be
\dt = \mbox{const.}g^{\frac{1}{4}}(x'') g^{-\frac{1}{4}}(x')\; .
\label{eq76}
\ee
Specializing to the Euclidean case, i.e. $g_{\m \n}=\d_{\m \n}$,
the ``const.'' is identified with ``1''. Furthermore, for the
{\em periodic boundary conditions} (\ref{eq80}), equation
(\ref{eq76}) reduces to
\be
sdet(G^{+})=1 \hspace{.3in}(\mbox{for:}\hspace{3mm} x''=x') \; .
\label{eq77}
\ee
Equation (\ref{eq77}) is a direct consequence of supersymmetry. It results in
a great deal of simplifications in (\ref{eq66}), particularly in
the higher-loop calculations, \cite{R-factor}.
\section{The Derivation of the Index of Dirac Operator
(A Proof of Theorem 1)}
Combinning (\ref{eq11}),(\ref{eq54}),(\ref{eq64}),(\ref{eq66}),(\ref{eq67}) and
(\ref{eq77}), one obtains the index of $\not\!\partial$ in the form:
\newcommand{\bo}{\b \rightarrow 0}
\be
index(\not\!\partial )=\frac{1}{(2\pi i)^{l}}\int K(x,\p ;\bo )\: \Theta
\: d^{m}\!x\, d^{m}\!\p ,
\label{eq78}
\ee
where
\be
K(x,\p ;\bo ):=\br x,\p ;\bo \mid x,\p ;0\kt \stackrel{
\mbox{\footnotesize WKB}}{=} Z c
e^{i\so } \ll sdet(G)\rr^{\frac{1}{2}} ,
\label{eq79}
\ee
and $\Theta $ is the superjacobian associated with the change of variables
of integration from $\xi ,\xi^{*}$'s to $\p$'s.

In (\ref{eq79}), the periodic boundary conditions must be adapted,
i.e.
\be
\begin{array}{ccccc}
x(0)&=&x(\b )&=:&\xo \\
\p (0)&=&\p (\b )&=:&\po \; .
\end{array}
\label{eq80}
\ee
This is consistent with supersymmetry (\ref{eq20}),
\cite{cecotti,ag3,goodman}.

Requiring (\ref{eq80}) and considering $\b \rightarrow 0$, the only
solution of the classical dynamical equations (\ref{eq22}), with
$\k =0$, is the constant configuration:
\be
\xo (t) = \xo \hspace{.4in} \po (t) = \po \; .
\label{eq81}
\ee
Substituting (\ref{eq81}) in (\ref{eq17}), one has \footnote{One must
take note of the end point contribution to $\so$. For details see
the Appendix.}:
\be
\so = 0 \; .
\label{eq81.1}
\ee
Thus, the factor $e^{i\so}$ in (\ref{eq79}) drops. Another important
consequence of (\ref{eq81}) is that unlike $ _{i,}S_{,j}$, (\ref{eq24}),
 $ _{i,}S_{0,j}$ are tensorial quantitities.
This allows one to work with normal coordinates, \cite{cd1}, centered at
$\xo$, in which
\bea
g_{0\m \n}& :=& g_{\m \n}(\xo ) = \d_{\m \n} \label{eq83} \\
g_{\m \n ,\s}(\xo )&=&\G^{\s}_{\m \n}(\xo ) = 0 \label{eq84} .
\eea
In the rest of this section, all the fields are evaluated in such a
coordinate system. The final results hold true for arbitrary coordinates
since the quantities of interest are tensorial.

Substituting (\ref{eq81}) in  (\ref{eq24}), using (\ref{eq84}),
and noting that all the $C$'s vanish and one obtains:
\newcommand{\cur}{{\cal R}}
\bea
_{\t ,}S_{0,\z '} & = & \ll -g_{0\t \z }\ppt + \cur_{\t \z}\pt \rr
\d (t-t')  \nn  \\
_{\t ,}S_{0,\g '} & = & _{\l ,}S_{0,\z '}  = 0  \label{eq87}  \\
_{\l ,}S_{0,\g '} & = & \ll ig_{0 \l \g}\pt \rr \d (t-t') .\nn
\eea
Here, $g_{0 \t \z}=\d_{\t \z}$ are retained for convenience, and
\be
\cur_{\t \z} := \mbox{\fs$\frac{i}{2}$}R_{\d \th \t \z}(\xo )\po^{\d}\po^{\th}.
\label{eq88}
\ee
The Green's functions (\ref{eq68}) are also tensorial quantities.
They can be explicitly computed:
\bea
G^{\z \xi '} & = & g_{0}^{\z \m} \ll \th (t-t')\frac{(e^{\cur (t-\b )}-1)
(1-e^{-\cur t'})}{\cur (1-e^{-\cur \b} )} + \right. \label{eq90} \\
& & \hspace{.5in}\left. + \th (t'-t)\frac{(e^{\cur t}-1)
(1-e^{-\cur (t'-\b )})}{\cur (e^{\cur \b}-1)} \rr_{\m}^{\xi} \nn \\
G^{\z \d '} & = & 0   \label{eq91} \\
G^{\g \xi '} & = & 0 \label{eq91.1} \\
G^{\g \d '} & = & \frac{i}{2}g_{0}^{\g \d}\ll \th (t-t') - \th (t'-t) \rr\; .
\label{eq92}
\eea
In (\ref{eq90}), the expression inside the bracket is to be interpreted
as a power series in:
\be
\cur := (\cur_{\t}^{\m}) := ( g_{0}^{\m \n}\cur_{\t \n}).
\label{eq89}
\ee
This is a finite series due to the presence of $\po$'s in (\ref{eq88})
\footnote{One can use all the properties of the ``$\exp$'' and other
analytic functions with arguments such as $\cur $. The only
rule is that the power series expansion must be postponed until
all other operations are performed.}.
Equation (\ref{eq90}) is obtained starting from the following
ansatz:
\newcommand{\kp}{{\cal X}_{+}}
\newcommand{\km}{{\cal X}_{-}}
\newcommand{\kpm}{{\cal X}_{+\m}^{\xi}}
\newcommand{\kmm}{{\cal X}_{-\m}^{\xi}}
\be
G^{\z \xi '} = \ll \th (t-t') (\frac{t}{\b}-1) t' g^{\z \m}
\kpm (t,t') + \th (t'-t) (\frac{t'}{\b}-1) t\, g^{\z \m}
\kmm (t,t') \rr.
\label{eq93}
\ee
(\ref{eq93}) satisfies the boundary conditions (\ref{eq68.1}), i.e.
\be
G^{\z \xi '} = 0 \hspace{.5in}\mbox{if $t$ or $t'$ $=$ $0$ or $\b$} .
\label{eq90.1}
\ee
Moreover, the reduction to $\p =0$ case is equivalent to choosing
${\cal X}_{\pm}=\I $. Imposing (\ref{eq68}) on (\ref{eq93})
and using (\ref{eq87}), one obtaines the following equations:
\bea
t>t' & : & (t-\b )\ll \ppt \kp -\cur \pt \kp \rr + 2\ll \pt \kp
-\mbox{\fs$\frac{1}{2}$} \cur \kp \rr =  0 \label{eq94} \\
t<t' & : & t \ll \ppt\km -\cur \pt \km \rr + 2\ll \pt \km
-\mbox{\fs$\frac{1}{2}$} \cur \km \rr =  0 \label{eq95} \\
t=t' & : & \ll \frac{t}{\b}\kp -(\frac{t}{\b}-1)\km + \right.
\label{eq96}  \\
& & \left.+ t(\frac{t}{\b}
-1) \left( \pt \kp - \pt \km - \cur \kp + \cur \km \right) \rr_{t=t'}
= \I . \nn
\eea
(\ref{eq94}) and (\ref{eq95}) are easily solved by the power series method.
The final result, (\ref{eq90}), follows using (\ref{eq68.2}) and (\ref{eq96}).
The derivation of (\ref{eq91}),(\ref{eq91.1}), and (\ref{eq92}) is
straightforward.

The next step is to compute $sdet(G)$. This is accomplished by
considering a functional variation in the metric tensor, $\d g_{\m \n}$,
and using the definition of $sdet$, (\ref{eq69}). After considerable
amount of algebra and repeated use of the symmetries of $g_{\m \n}$,
and $\cur_{\m}^{\n}$, one arrives at the following expression:
\be
\d \ln [sdet(G)] = \th (0) tr\ll 2\d \ln (\b \cur ) + \b \d \cur
(\frac{1+e^{\b \cur}}{1-e^{\b \cur}}) \rr\; .
\label{eq97}
\ee
Setting $\th (0)=1/2$ and integrating the right hand side of (\ref{eq97}),
one has:
\be
\d \ln [sdet(G)] = \d tr \left( \ln \ll \frac{-\frac{\b \cur}{2}}{\sinh
(\frac{\b \cur}{2})} \rr \right)\; .
\label{eq98}
\ee
Since $\cur$ is  antisymmetric in its indices, it can be put in the
following block-diagonal form:
\be
\cur = diag \left( \ll
\begin{array}{cc}
0 & \cur_{i} \\
-\cur_{i} & 0
\end{array}
\rr : i=1,\cdots ,l \right) \; .
\label{eq99}
\ee
\newcommand{\AO}{\frac{\frac{\b \cur}{2}}{\sinh (\frac{\b \cur}{2})} }
Another important observation is that $\AO$ and hence $\ln [-\AO ]$
are polynomials in $(\frac{\b \cur}{2})^{2}$ . In view of (\ref{eq99}),
it is easy to see that $\cur^{2}$ is diagonal, namley
\be
\cur^{2}=diag(-\cur_{1}^{2},-\cur_{1}^{2},\cdots,-\cur_{l}^{2},-\cur_{l}^{2}).
\label{eq100}
\ee
Implementing (\ref{eq100}) in (\ref{eq98}), one has:
\renewcommand{\cur}{i{\cal R}_{j}}
\newcommand{\pl}{\prod_{j=1}^{l}}
 \[ \d \ln [sdet(G)] = 2 \d \ln \pl \ll -\AO \rr  , \]
and finally
\be
sdet(G)^{\frac{1}{2}} = \tilde{c} \pl \ll \AO \rr ,
\label{eq101}
\ee
where $\tilde{c}$ is a constant of functional integration.
Substituting (\ref{eq101}) and (\ref{eq81}) in (\ref{eq79}), one obtains
the kernel in the form:
\be
K(x,\p ;\bo ) =  Zc\,\tilde{c}\pl \ll \AO \rr .
\label{eq102}
\ee
\renewcommand{\cur}{{\cal R}}
\newcommand{\crj}{\cur_{j}}
\newcommand{\oj}{\Omega_{j\d \th }\po^{\d}\po^{\th}}
\newcommand{\boj}{\frac{\b \oj}{2}}
To write $\crj$ as functions of $\xo$ and $\po$, one needs to also
block-diagonalize:
\be
\left( \mbox{\fs$\frac{1}{2}$}R_{\m \n \d \th}
\mbox{\fs$(\xo)$}\,\po^{\d}\po^{\th} \right) = diag
\left( \ll
\begin{array}{cc}
0 & \oj \\
-\oj & 0
\end{array} \rr : \mbox{\fs$j=1,\cdots ,l$} \right) .
\label{eq103}
\ee
Combinning (\ref{eq88}),(\ref{eq89}),(\ref{eq99}),(\ref{eq102}) and
(\ref{eq103}), one is led to
\be
K(\xo ,\po ;\bo )= Zc\tilde{c} \pl \ll \frac{\boj}{\sinh (\boj )}
\rr .
\label{eq104}
\ee
The proportionality constant $Zc\,\tilde{c}=:\tilde{Z}$ is determined
by specializing to the $M=\R^{m}$ case. Using the results of
\cite[\S 5,\S 6]{bd1}, one has
\be
\tilde{Z} = (2\pi i dt)^{-\frac{\b m}{2 dt}}\times (2\pi i)^{
-\frac{\b l}{dt}+l}\; .
\label{eq105}
\ee
The limit $\b \rightarrow 0$ is taken by setting
\be
\b =dt\; .
\label{eq106}
\ee
Finally, combinning (\ref{eq78},\ref{eq104}) and (\ref{eq106}), and realizing
that $e^{\m}_{i}(\xo)=\d^{\m}_{i}$ so that $\Theta =i^{-l}$, one has
\be
index(\not\!\partial )=\frac{1}{(2\pi )^{l}}\int \pl \ll
\frac{\boj}{\sinh (\boj )}\rr \frac{d^{m}\po \, d^{m}\!\xo}{(2\pi i\b )^{l}}.
\label{eq108}
\ee
The $\po$-integration rules, \cite{bd1}:
\be
\begin{array}{ccc}
\int d\po &=& 0 \\
\int \po \, d\!\po &=& \sqrt{2\pi i}
\end{array}
\label{eq109}
\ee
allow only the highest degree term in the integrand to survive. This
implies the cancellation of $\b $'s. Performing the $\po$-integrations
yields an expression for (\ref{eq108}) which is identical with the
following:
\bea
index(\not\!\partial )&=&\frac{1}{(2\pi )^{l}}\int_{M}\pl \ll
\frac{\frac{\Omega_{j}}{2}}{\sinh (\frac{\Omega_{j}}{2})}
\rr_{\mbox{top}} \nn \\
& = & \int_{M}\pl \ll \frac{\frac{\Omega_{j}}{4\pi}}{\sinh
(\frac{\Omega_{j}}{4\pi})} \rr_{\mbox{top}}\; .
\label{eq110}
\eea
In (\ref{eq110}), $\Omega_{j}$ are defined by (\ref{eq5}) and the identity
\[ \Omega_{j} = \Omega_{j\d \th}\: dx^{\d}\wedge dx^{\th}. \]
(\ref{eq110}) is precisely the statement of Theorem 1, i.e. (\ref{eq4}).
The fact that the integrand in (\ref{eq110})
is an even polynomial in $\frac{\Omega{_j}}{4\pi}$ implies that only
for $l=2k$ , i.e. $m=4k$, is the index nonvanishing.

\section{The Derivation of the Index of the Twisted Dirac Operator
(A Proof of Theorem 2)}

Equations (\ref{eq59}),(\ref{eq62}),(\ref{eq63}),(\ref{eq64}),
(\ref{eq66}),(\ref{eq67}), and (\ref{eq77}) provide the following formula
for the index
\be
index(\not\!\partial_{V})=\left. \frac{\partial}{\partial \l}
\right|_{\l =0} \ll
\frac{1}{(2\pi i)^{l+n}}\int K(x,\p ,\e ;\b\rightarrow 0)\,\Theta\,
d^{m}\! x\, d^{m}\!\p\, d^{n}\!\e^{*}\, d^{n}\!\e \rr
\label{eq111}
\ee
where
\be
K(x,\p,\e ;\b\rightarrow 0):=\br x,\p,\e^{*};\b\rightarrow 0 \mid
x,\p ,\e ;0 \kt\stackrel{\mbox{\footnotesize WKB}}{=\hspace{-1.5mm}=}
Z'c'e^{i\so}[sdet(G)]^{\frac{1}{2}} .
\label{eq112}
\ee
The first step in the calculation of the kernel is to obtain the
classical solution of the dynamical equations, (\ref{eq22}), in the
$\b\rightarrow 0$ limit.

As one can see from the analysis of Section 6, absorbing a factor
of $\sqrt{\b}$ in $\p$'s can take care of the limiting process
automatically. Furthermore, following \cite{manes-zumino}, define
the parameter:
\[ s:=\frac{t}{\b}, \]
and expand the coordinate variables in powers of $\b$:
\bea
x(t)&=&\tilde{x}_{0}(s)+\tilde{x}_{1}(s)\b +O(\b^{2}) \nn \\
\p (t)&=&\frac{1}{\sqrt{\b}}\ll \tilde{\p}_{0}(s)+\tilde{\p}_{1}(s)
\b +O(\b^{2})\rr \label{eq113} \\
\e (t)&=&\tilde{\e}_{0}(s)+\tilde{\e}_{1}(s)\b +O(\b^{2}). \nn \\
\e^{*}(t)&=&\tilde{\e}^{*}_{0}(s)+\tilde{\e}^{*}_{1}(s)\b+O(\b^{2}) \; .\nn
\eea
The appropriate boundary conditions for (\ref{eq112}) are the periodic
boundary conditions \footnote{In fact, one can adopt antiperiodic boundary
conditions for $\e$'s and $\e^{*}$'s, \cite{ag3}. Since the relevant
trace is taken over the 1-$\e$-particle state vectors, this would
introduce an extra minus sign.}, i.e., (\ref{eq80}) and:
\bea
\e & = &\e ':=\e (t=0) =\e (\b ) =: \e '' \nn \\ \nopagebreak[3]
\e^{*'}& := &\e^{*}(t=0)=\e^{*}(t=\b )=:\e^{*''}=\e^{*} \; .
\label{eq114}
\eea

It follows from (\ref{eq80}),(\ref{eq114}), and (\ref{eq113}) that the
dynamical
equations are solved by:
\bea
\xo (t)&=&\xo + O(\b ) \nn \\
\po (t)&=&\frac{1}{\sqrt{\b}}\ll \tilde{\p}_{0} + \tilde{\p}_{1}(s) \b
+O(\b^{2}) \rr \label{eq115} \\
\e_{0} (t)&=&\tilde{\e}_{0}(s)+O(\b ) \nn \\
\e_{0}^{*}(t)&=&\tilde{\e}^{*}_{0}(s)+O(\b ), \nn
\eea
where $\xo$ and $\po =:\frac{1}{\mbox{\fs$\sqrt{\b}$}}\tilde{\p}_{0}$
are constants.
Adopting a normal coordinate frame centered at $\xo$ and using (\ref{eq83}),
(\ref{eq84}) and
\[ A_{\m}^{ab}(\xo)=0 ,\]
one has:
\newcommand{\te}{\tilde{\e}_{0}}
\newcommand{\cuf}{{\cal F}}
\bea
\frac{d}{ds}\tilde{\p}_{1}^{\g}-ig_{0}^{\g \l}F_{\l \d}^{ab}(\xo )
\tilde{\p}_{0}^{\d}\tilde{\e}_{0}^{a*}\tilde{\e}_{0}^{b}&=&0 \nn \\
\frac{d}{ds}\tilde{\e}_{0}^{a}-i(\tcuf^{ab}+\a \d^{ab})\te^{b}&=&0
\label{eq116} \\
\frac{d}{ds}\te^{a*}+i(\tcuf^{ba}+\a \d^{ba})\te^{b*}&=&0 .\nn
\eea
Here $\tcuf$ is defined by
\be
\tcuf = \left( \tcuf_{ab} \right) := \left( \mbox{\fs$\frac{1}{2}$}
F_{\g \l}^{ab}(\xo)\, \tilde{\p}_{0}^{\g}\tilde{\p}_{0}^{\l} \right)\; .
\label{eq117}
\ee
The quantity:
\be
\cuf =\left( \cuf_{ab} \right) := \left( \mbox{\fs$\frac{1}{2}$}
F_{\g \l}^{ab}(\xo )\, \po^{\g}\po^{\l} \right) = \mbox{\fs$\frac{1}{\b}$}
\tcuf
\label{eq118}
\ee
will also be used.

The next step is to compute $\so$. This is done in the Appendix. The
final result is :
\be
\so = -i\e^{a*}\ll e^{i(\b \cuf +\a \In )}-\In \rr^{ab}\e^{b}\; .
\label{eq119}
\ee

Next, one needs to calculate the Feynman propagator $G$. This is
done in two steps. First, consider the special case of
\bea
\xo (t)&=&\xo \nn \\
\po (t)&=&\po = \mbox{\fs$\frac{1}{\sqrt{\b}}$}\tilde{\p}_{0} \label{eq120} \\
\e_{0}(t)&=&0 \nn \\
\e^{*}_{0}(t)&=&0 \; . \nn
\eea
It is clear that (\ref{eq120}) satisfies the dynamical equations,
(\ref{eq22}). Substituting (\ref{eq120}) in (\ref{eq24}) and (\ref{eq25}),
one recovers (\ref{eq87}). The other nonvanishing $_{i,}S_{0,j}$'s are:
\be
\begin{array}{ccc}
_{b,}S_{0,a^{*'}}&=&\ll i\d_{ba}\pt -(\cuf_{ab}+\frac{\a}{\b}\d_{ba})\rr
\d (t-t') \\
_{b^{*},}S_{0,a'}&=&\ll i\d_{ba}\pt +(\cuf_{ba}+\frac{\a}{\b}\d_{ba})\rr
\d (t-t')\; .
\end{array}
\label{eq121}
\ee
In other words, one has:
\be
\left( \, _{i,}S_{0,j'} \right)=\ll
\begin{array}{cccc}
_{\t ,}S_{0,\z '} & 0 & 0 & 0 \\
0 & _{\l ,}S_{0,\g '} & 0 & 0 \\
0 & 0 & 0 & _{b,}S_{0,a^{*'}} \\
0 & 0 & _{b^{*},}S_{0,a'} & 0
\end{array} \rr\; .
\label{eq122}
\ee
(\ref{eq122}) suggests:
\be
\left( G^{ij'} \right) = \ll
\begin{array}{cccc}
G^{\z \xi '} & 0 & 0 & 0 \\
0 & G^{\g \e '} & 0 & 0 \\
0 & 0 & 0 & G^{ac^{*'}} \\
0 & 0 & G^{a^{*}c'} & 0
\end{array} \rr\; .
\label{eq123}
\ee
In view of (\ref{eq68}), it is clear that $G^{\z \xi '}$ and $G^{\g \e '}$
are given by equations (\ref{eq90}) and (\ref{eq92}), respectively. Defining
\[ G_{1}(t,t') = \left( G_{1}^{ac}(t,t') \right) := \left( G^{a^{*}c'}\right)\]
\[ G_{2}(t,t') = \left( G_{2}^{ac}(t,t') \right) := \left(
G^{ac^{*'}}\right),\]
and using (\ref{eq123}),(\ref{eq87}),(\ref{eq121}),(\ref{eq90}),(\ref{eq92}),
and (\ref{eq68}), one obtains:
\be
\begin{array}{ccc}
\ll i\pt - (\cuf^{*}+\frac{\a}{\b}) \rr G_{1}(t,t')&=&-\d (t-t') \\
\ll i\pt + (\cuf +\frac{\a}{\b}) \rr G_{2}(t,t')&=&-\d (t-t') \; .
\end{array}
\label{eq124}
\ee
In (\ref{eq124}), use has been made of the fact that $\cuf$ is hermitian
and hence
\[ \cuf^{\rm transpose}=\cuf^{*}. \]
To compute $G_{1}$ and $G_{2}$, consider the ansatz:
\newcommand{\xx}{{\cal X}}
\newcommand{\yy}{{\cal Y}}
\be
G_{k}(t,t')=\th (t-t')e^{i\xx_{k}(t,t')}-\th (t'-t)e^{i\yy_{k}(t,t')}
\hspace{12mm}(k=1,2) .
\label{eq125}
\ee
Substituting (\ref{eq125}) in (\ref{eq124}), one obtains:
\bea
t>t' &:&\pt\xx_{1}=-(\cuf^{*}+\frac{\a}{\b})\hspace{2mm},\hspace{2mm}
\pt\xx_{2}=\cuf + \frac{\a}{\b} \nn \\
t<t' &:&\pt\yy_{1}=-(\cuf^{*}+\frac{\a}{\b})\hspace{2mm},\hspace{2mm}
\pt\yy_{2}=\cuf + \frac{\a}{\b} \nn \\
t=t' &:&\ll e^{i\xx_{k}}+e^{i\yy_{k}}\rr_{t=t'}=i \hspace{5mm}k=1,2. \nn
\eea
Finally, using (\ref{eq68.2}):
\[ G_{1}(t,t')=-G_{2}^{\rm transpose}(t',t) ,\]
one arrives at the following expressions:
\be
\begin{array}{ccccc}
G_{1}(t,t')&=&\left( G^{a^{*}c'} \right)&=&\frac{i}{2}
e^{-i(\cuf^{*}+\frac{\a}{\b})(t-t')}\ll \th (t-t')-\th (t'-t) \rr \\
G_{2}(t',t)&=&\left( G^{ac^{*'}} \right)&=&\frac{i}{2}
e^{i(\cuf +\frac{\a}{\b})(t-t')}\ll \th (t-t')-\th (t'-t) \rr\; .
\end{array}
\label{eq126}
\ee
Next step is to observe that, in the limit: $\b \rightarrow 0$, (\ref{eq123})
with (\ref{eq90}),(\ref{eq92}), and (\ref{eq126}) actually satisfies equation
(\ref{eq68}) even in the general case of (\ref{eq115}). To see this, it is
sufficient to examine:
\bea
_{\t ,}S_{0,\z '}&=&\b^{-3}\ll -g_{0\t \z}\frac{\partial^{2}}{\partial s^{2}}
+\frac{i}{2}R_{\l \g \z \t}(\xo )\, \tilde{\p}_{0}^{\g}\tilde{\p}_{0}^{\l}\pss
+O(\b )\rr \d (s-s') \label{eq127} \\
_{\l ,}S_{0,\g '}&=&\b^{-2}\ll ig_{0\l \g}\pss +O(\b )\rr\d (s-s')
\label{eq128}
\eea
and
\bea
_{b,}S_{0,a^{*'}} & = &\b^{-2}\ll i\d_{ba}\pss -(\tcuf_{ab}+\a\d_{ab})+O(\b
)\rr
\d (s-s') \nn \\
_{b^{*},}S_{0,a'} & = & \b^{-2}\ll i\d_{ba}\pss +(\tcuf_{ba}+\a\d_{ba})+O(\b )
\rr \d (s-s') .
\label{eq129}
\eea
The next step is to compute  $sdet(G)$. Since $G$ is block-diagonal, one
has:
\be
sdet(G)=sdet(G_{0}).sdet(G_{\e})
\label{eq130}
\ee
where
\[ G_{0} := \left( \begin{array}{cc}
G^{\z\xi '} & 0 \\
0 & G^{\g\l '} \end{array} \right) \mbox{\hspace{2mm}and\hspace{2mm}}
G_{\e} := \left( \begin{array}{cc}
0 & G^{ac^{*'}} \\
G^{a^{*}c'} & 0 \end{array} \right) . \]
Clearly, $sdet(G_{0})$ is given by (\ref{eq101}). $sdet(G_{\e})$ is
computed following the procedure of Section 6. Applying (\ref{eq69})
and (\ref{eq71}) to $G_{\e}$ and writing only the nonvanishing terms,
one has:
\be
\d \ln sdet(G_{\e}) = \int_{0}^{\b}dt\int_{0}^{\b}dt' \ll
-\, _{c,}\d\! S_{0,a^{*'}}G^{a^{*'}c}-\, _{c^{*},}\d\! S_{0,a'}G^{a'c^{*}} \rr
{}.
\label{eq131}
\ee
Let us define:
\newcommand{\zz}{{\cal Z}}
\be
\zz (t,t'') := \int_{0}^{\b}dt'\ll -\, _{c,}\d\! S_{0,a^{*'}}G^{a^{*'}c''}
-\, _{c^{*},}\d\! S_{0,a'}G^{a'c^{*''}} \rr .
\label{eq132}
\ee
Taking the functional variation of (\ref{eq129}) with respect to
$\d\cuf$ and substituting the result in (\ref{eq132}), one has:
\bea
\zz (t,t'') & = & -\d\cuf_{ac}G^{a^{*}c''}+\d\cuf_{ca}G^{ac^{*''}}
\label{eq133} \\
& = & tr\ll -\d\cuf^{*}\, G_{1}(t,t'')+\d\cuf\, G_{2}(t,t'')\rr \nn \\
& = & tr \ll \d e^{-i(\cuf^{*}+\frac{\a}{\b})(t-t'')}+
             \d e^{i(\cuf +\frac{\a}{\b})(t-t'')} \rr \mbox{\fs$
\ll \frac{\th (t-t'') -\th (t''-t)}{2(t''-t)} \rr$} \nn \\
& = & tr \ll \d \left( \cos [ (\cuf +\frac{\a}{\b})(t-t'')] \right) \rr
\mbox{\fs$\ll \frac{\th (t-t'')-\th (t''-t)}{t''-t} \rr$} \nn \\
& = & \d tr \ll \frac{1}{2}(\cuf +\frac{\a}{\b})^{2}(t''-t)+O(t-t'')^{3} \rr
\ll \th (t-t'') -\th (t''-t) \rr. \nn
\eea
Combinning equations (\ref{eq131}),(\ref{eq132}), and (\ref{eq133}), one
obtains:
\[ \d \ln sdet(G_{\e}) = \int_{0}^{\b}dt \:\zz (t,t) = 0 ,\]
and hence:
\be
sdet(G_{\e}) = const.
\label{eq134}
\ee
Equations (\ref{eq101}),(\ref{eq130}), and (\ref{eq134}) yeild:
\renewcommand{\cur}{i{\cal R}_{j}}
\be
\ll sdet(G)\rr^{\frac{1}{2}}=\tilde{c}'\pl \ll \AO \rr\; .
\label{eq135}
\ee

Substituting (\ref{eq119}) and (\ref{eq135}) in (\ref{eq112}), one
finds:
\be
K(\xo ,\po ,\e ;\b\rightarrow 0) = Z'c'\tilde{c}'\pl \ll \AO \rr
\mbox{\fs$\exp \left( \e^{a*}\ll e^{i(\b\cuf +\a\In)}-\In\rr^{ab}\e^{b}
\right)$}\; .
\label{eq136}
\ee
The constant $\tilde{Z}':=Z'c'\tilde{c}'$ is fixed by specializing to
the case of $\cuf =\a =0$. In this case (\ref{eq136}) must reduce to
(\ref{eq104}). Thus,
\[ \tilde{Z}'=\tilde{Z}=(2\pi i\b )^{-l}. \]
Substituting (\ref{eq136}) in (\ref{eq111}) and performing the $\e^{*}$
and $\e$ integrations, one has:
\be
index(\not\!\partial_{V})=\left. \frac{\partial}{\partial \l}
\right|_{\l =0} \left( \frac{1}{(2\pi )^{l}}\int \pl \ll \AO \rr
\mbox{\fs$ det \ll e^{i(\b\cuf +\a\In )} - \In \rr $}\right)  \; .
\label{eq137}
\ee
At this stage, one can take the $\l$-derivative. Since $\cuf$ is hermitian
it can be diagonalized, i.e.
\[ \cuf =: diag ( \cuf_{1},\cdots ,\cuf_{n}). \]
\newpage
Then,
\newcommand{\plo}{\left. \frac{\partial}{\partial \l}\right|_{\l =0}}
\bea
\plo det\ll e^{i(\b\cuf +\a\In)}-\In\rr & = & \plo \ll \prod_{a=1}^{n}
(\l e^{i\b\cuf_{a}}-1) \rr \nn \\
&=& (-1)^{n}\plo \ln \ll \prod_{a=1}^{n}(\l e^{i\b\cuf_{a}}-1) \rr \nn \\
& = & \sum_{a=1}^{n}e^{i\b\cuf_{a}} \nn \\
& = & tr\, (e^{i\b\cuf}) \; . \label{eq138}
\eea
Using (\ref{eq138}) and performing $\po$-integrations, one finally
arrives at:
\bea
index(\not\!\partial_{V})&=& \frac{1}{(2\pi )^{l}}\int_{M}
\ll \pl \left( \frac{\frac{\Omega_{j}}{2}}{\sinh (\frac{\Omega_{j}}{2})}
\right) tr\, (e^{iF})\rr_{\rm top.} \nn \\
& = & \int_{M}\ll \pl \left( \frac{\frac{\Omega_{j}}{4\pi}}{
\sinh (\frac{\Omega_{j}}{4\pi})} \right) tr\, (e^{\frac{iF}{2\pi}})
\rr_{\rm top.} .
\label{eq139}
\eea
In (\ref{eq139}), $\Omega_{j}$ and $F$ are defined by (\ref{eq5}) and
(\ref{eq9.1}), respectively. This completes the proof of Theorem 2.

\section{Remarks and Discussion}
The following is a list of remarks concerning some of the aspects of
the present work:
\begin{itemize}
\begin{enumerate}
\item The reality condition on the Lagrangian (\ref{eq17}), requires
$F_{\m\n}$ to be antihermitian matrices. This is equivalent to
requiring the structure group of the bundle $V$ to be (a subgroup)
of $U(n)$. This is consistent with the existence of a hermitian
structure on $V$. For, any hermitian vector bundle can be reduced
to one with $U(n)$ as its structure group.
\item The term $i\e^{a*}\dot{\e}^{a}$ in the Lagrangian (\ref{eq17})
can be replaced with
\[ \mbox{\fs$\frac{i}{2}$} (\e^{a*}\dot{\e}^{a}-\dot{a}^{a*}\e^{a}) \]
with no consequences for the action. This is because the boundary
conditions on $\e$'s and $\e^{*}$'s are periodic.
\item In the Peierls quantization scheme, the momenta, ``$p_{\m}$'',
are not a priori fixed. They may be chosen in a way that facilitates
the analysis of the problem. In particular, the choice: (\ref{eq31})
has obvious advantages. The factor ordering problem may arise in
general. In the case of th index problem, however, the requirement that
the supersymmetry generator be identified with the elliptic
operator in question, enables one to choose the appropriate ordering.
It is also remarkable that in the special case that the structure
group of $V$ can be chosen $SU(n)$, the factor ordering ambiguities,
(\ref{eq33.1}), disappear. It seems that the obstruction for having
a unique quantum system is the first Chern class.
\item The problem of nonuniqueness of the momentum operators was
not addressed in Section~4. In general,
\[ {\bf p}:=p_{\m}\, dx^{\m} \]
is unique up to closed forms, $\omega\in H^{1}(M,\R\! )$, \cite{bd1}.
This is reflected in the path integral formulation as the
necessity of concidering different homotopy meshes of paths. This
is, however, not relevant to the index problem. In general, the
index of an elliptic operator is a map:
\[ index\, :\:K(TM)\,\rightarrow \Z \; , \]
where $K(TM)\cong K(M)\cong\bigoplus_{i} H^{2i}(M,\!\Z )$ is the
Grothendieck's K-group (ring), \cite{shanahan}. Thus the index does
not detect the first cohomology group of M.
\item The procedure used in the evaluation of the index in this paper
are essentially based on a single basic assumption. This is the
Gaussian functional integral formula, (\ref{eq67}). The other
important ingredient is the definition of the superdeterminant,
(\ref{eq69}),
which is assumed to hold even for infinite dimensional matrices.
The latter can be easly shown to be a consequence of (\ref{eq67}).
\item The factor $[ sdet\, G^{+}]^{-\frac{1}{2}}$, in (\ref{eq64}), is
usually constant for the case of flat spaces. However, it contributes
in an essential way in the case of curved spaces. For the system
considered in this paper, it was explicitely calculated and shown
to be 1. It is not difficult to see that the amazing cancellations
are due to supersymmetry. Considering the complexity of the system,
(\ref{eq17}), (\ref{eq24}), and (\ref{eq25}), this might be a
general pattern for a large class of supersymmetric systems.
\item The appearence of $\frac{\hbar^{2}}{8}R$ in the Hamiltonian,
(\ref{eq35}), can be verified independently by comparing the heat
kernel and loop expansions of the path integral, (\ref{eq64}).
This  term contributes to the linear term in
the heat kernel expansion, \cite{bd1}. On the other hand, the
presence of $\hbar^{2}$ is reminiscent of the contribution
to the path integral in the 2-loop order. The complete
2-loop analysis of this problem is the subject of \cite{R-factor}.
\end{enumerate}
\end{itemize}

\section{Conclusion}
The Peierls quantzation scheme provides a systematic procedure
for quantizing superclassical systems. Supersymmetry results in
remarkable simplifications in the path integral evaluation
of the kernel. It also leads to a direct proof of the twisted
spin index theorem. The supersymmetric proof of the Atiyah-Singer
index theorem is based on the assumption that the ordinary
Gaussian integral formula holds even in the functional integral
case. The index theorem is one of the best established
mathematical results. Thus, the existence of its  supersymmetric proof
can be viewed as another sign of the validity and power of the
path integral techniques. In particular, it is remarkable to
observe that even the normalization constants agree with those
chosen by mathematicians.
\section*{Acknowledgements}
This project was suggested to the author by Prof. Bryce DeWitt.
He would like to thank Prof. DeWitt  for invaluable discussions and
his guidance and support. He would also like to acknowledge Prof.'s Luis Boya
and Cecile Dewitt-Morette for reading the first draft and for their
most helpful comments and suggestions.

\appendix
\section*{Appendix: Coherent State Path Integral and Calculation of $\so$}
In the coherent state path integral formula for the kernel, one
must also include the end point contributions to the action,
\cite{it-zu,faddeev}\footnote{In fact, the action function can be
shown to include the end point contributions automatically. This
was pointed out to the author by Cecile DeWitt.}.
This can be done implicitly by including
step functions at the end points of the paths. An example of this
is provided in \cite{bd1}.

\renewcommand{\z}{\zeta}
Consider a fermionic quantum system and let $\hat{\z}^{i\dagger}$ and
$\hat{\z}^{i}$ be the creation and annihilation operators:
\bea
\{\hat{\z}^{i\dagger},\hat{\z}^{j}\}&=&\d^{ij} \nn \\
\{\hat{\z}^{i},\hat{\z}^{j}\} & = & \{\hat{\z}^{i\dagger},\hat{\z}^{j\dagger}
\} = 0 \; . \nn
\eea
The coherent states are defined by:
\bea
\mid\!\z\kt &:=& e^{\frac{1}{2}\z^{i*}\z^{i}+\hat{\z}^{i\dagger}\z^{i}}
\mid\! 0\kt \nn \\
\br \z^{*}\!\mid &:=& \mid\! \z \kt^{\dagger}, \nn
\eea
where $\mid\! 0\kt$ is the vacuum and $\z \in \C$. Then one has:
\[ \hat{\z}^{i}\mid\! \z\kt =\z^{i}\mid\!\z\kt \hspace{5mm}\mbox{and}
\hspace{5mm}\br\z^{*}\!\mid \hat{\z}^{i\dagger} = \z^{i*}\br\z^{*}\!\mid \; .\]
In the path integral formula for the kernel:
\be
K(\z^{*''},t''\mid\z ',t'):=\br\z^{*''};t''\mid\z ';t'\kt =
\int e^{iS}(sdet\, G^{+})^{-\frac{1}{2}}{\cal D}\z^{*}\, {\cal D}\z\:,
\label{eqa1}
\ee
one must note that $\z^{*''}\neq (\z '')^{*}$. In fact a priori
only
\be
\z (t')=\z ' \hspace{1cm}\mbox{and}\hspace{1cm}\z^{*}(t'')=\z^{*''}
\label{eqa2}
\ee
are fixed. The boundary conditions on
\[ \z (t'')=\z ''\hspace{1cm}\mbox{and}\hspace{1cm}\z^{*}(t')=\z^{*'}\]
depend on the specifics of the problem.
In (\ref{eqa1}), the action is given by:
\be
S=\int_{\z ',t'}^{\z '',t''}dt\ll \frac{1}{2i}(\z^{i*}\dot{\z}^{i}
-\dot{\z}^{i*}\z^{i})-h(\z^{*},\z ,t)\rr
\label{eqa3}
\ee
where $h(\z^{*},\z ,t)$ is the normal symbol of the Hamiltonian,
\cite{it-zu}. The paths are given by:
\be
\begin{array}{ccc}
\z^{*}(t) & = & \lim_{\epsilon\rightarrow 0}\ll \th (t-t'-\epsilon )
\z_{c}^{*}(t) + \th (t'-t+\epsilon )\z^{*'}\rr \\
\z (t) & = & \lim_{\epsilon\rightarrow 0}\ll \th (t''-t-\epsilon )
\z_{c}(t) + \th (t-t''+\epsilon )\z ''\rr \; .
\end{array}
\label{eqa4}
\ee
In (\ref{eqa4}), $\z_{c}^{*}(t)$ and $\z_{c}(t)$ are paths connected
to the end points, i.e. :
\be
\z_{c}^{*}(t'')=\z^{*''}\hspace{1cm}\mbox{and}\hspace{1cm}
\z_{c}(t')=\z '
\label{eqa5}
\ee
In general,
\[ \z_{c}^{*}(t')\neq\z^{* '}\hspace{1cm}\mbox{and}\hspace{1cm}
\z_{c}(t'')\neq\z '' \: .\]
To compute $\so$, which appears in (\ref{eq112}), one needs to
specialize to the classical paths. In the limit $\b\rightarrow 0$,
these are given by (\ref{eq115}) and (\ref{eq116}).
Equations~(\ref{eq116}) are easily solved to give:
\bea
\tilde{\e}_{0c}^{a}(s) & = & \ll e^{i(\tcuf +\a\In )s} \rr^{ab}\e^{b'}
\label{eqa6} \\
\tilde{\e}^{a*}_{0c}(s) & = & \e^{b*''} \ll e^{i(\tcuf +\a\In )(1-s)}\rr^{ba}
\label{eqa7} \\
\tilde{\p}_{1c}^{\g}(s) & = & ig_{0}^{\g \l}F_{\l \d}^{ab}(\xo )
\tilde{\p}_{0}^{\d} E^{ab}(s) + C^{\g} \; , \label{eqa8}
\eea
where
\[ E^{ab}(s):=\int_{0}^{1} ds \, \tilde{\e}_{0c}^{a*}(s)\tilde{\e}_{0c}^{b}(s)
\, .\]
In particular, one has:
\be
\mbox{\fs$\frac{1}{2}$}g_{0\l\g}\tilde{\p}_{0}^{\l}\tilde{\p}_{1c}^{\g}(s)
=i\e^{a*''}\ll e^{i(\tcuf +\a\In )}\tcuf\rr^{ab}\e^{b'}s+C\; .
\label{eqa9}
\ee
In (\ref{eqa8}) and (\ref{eqa9}) $C^{\l}$ and $C$ are constants to be
determined by the boundary conditions on $\p$'s.

In view of equations (\ref{eq115}) and (\ref{eq17}), one has
\bea
\so &=& \int_{0}^{1}ds\ll \mbox{\fs$\frac{i}{2}$}g_{0\l\g}\,\tilde{\p}_{0}^{\l}
\,\dot{\tilde{\p}}_{1}^{\g}(s)+\mbox{\fs$\frac{i}{2}$} \left(
\tilde{\e}_{0}^{a*}(s)\,\dot{\tilde{\e}}_{0}^{a}(s)
-\dot{\tilde{\e}}_{0}^{a*}(s)\,\tilde{\e}_{0}^{a}(s) \right) + \right. \nn \\
& & \hspace{1.5cm} \left. + \tcuf^{ab}\,\tilde{\e}_{0}^{a*}(s)\,
\tilde{\e}_{0}^{b}(s) \rr + O(\b )\; .
\label{eqa10}
\eea
Defining
\[ \tilde{\xi}^{i}:=\sqrt{\b}\xi^{i}\; ,\]
and using (\ref{eq36}) and (\ref{eq38}), one may express
the first term on the right hand side of (\ref{eqa10}) in the form:
\newcommand{\tx}{\tilde{\xi}}
\newcommand{\dtx}{\dot{\tx}}
\newcommand{\iot}{\mbox{\fs$\frac{i}{2}$}}
\newcommand{\tp}{\tilde{\p}}
\bea
\int_{0}^{1} ds \ll \iot g_{0\l\g}\, \tp_{0}^{\l}\, \dot{\tp}_{1}^{\g}
(s) \rr &=& \int_{0}^{\b} dt \ll \iot
g_{0\l\g}\, \p_{0}^{\l}(t)\, \dot{\p}_{0}^{\g}(t) \rr + O(\b ) \label{eqa11} \\
&=& \int_{0}^{\b} dt \ll \iot \left( \xi_{0}^{i*}(t)\, \dot{\xi}_{0}^{i}(t)
-\dot{\xi}_{0}^{i*}(t)\, \xi_{0}^{i}(t) \right) \rr + O(\b )\nn \\
&=&
\int_{0}^{1} ds \ll \iot \left( \tx_{0}^{i*}(s)\, \dtx_{0}^{i}(s)-
\dtx_{0}^{i*}(s) \, \tx_{0}^{i}(s) \right) \rr + O(\b ) \; . \nn
\eea
In view of (\ref{eqa11}), it is clear that (\ref{eqa10}) is already in
the form demanded by (\ref{eqa3}).
Next step is to evaluate (\ref{eqa10}). Let us define:
\bea
I_{1}&:=&\int_{0}^{1}ds \ll \iot g_{0\l\g}\tp_{0}^{\l}\dot{\tp}_{1}^{\g}(s)
\rr \label{eqa12} \\
&\stackrel{\b\rightarrow 0}{=}&\int_{0}^{1}ds\ll \iot \left(
\tx_{0}^{i*}(s)\, \dtx_{0}^{i}(s) - \dtx_{0}^{i*}(s)\, \tx_{0}^{i}(s)
\right) \rr \nn \\
I_{2}&:=&\int_{0}^{1}ds\ll \iot\left( \tilde{\e}_{0}^{a*}(s)\,
\dot{\tilde{\e}}_{0}^{a}(s)-\dot{\tilde{\e}}_{0}^{a*}(s)\,
\tilde{\e}_{0}^{a}(s)\right)+\tcuf^{ab}\tilde{\e}_{0}^{a*}(s)\,
\tilde{\e}_{0}^{b}(s) \rr \; ,\label{eqa13}
\eea
so that
\be
\so = I_{1}+I_{2} \; .
\label{eqa14}
\ee
Replacing $\z$'s by $\tx_{0}$'s and setting $t'=0$ and $t''=1$ in
(\ref{eqa4}), it is a matter of simple algebra to show that:
\bea
I_{1}&=& I_{1c}+\partial \nn \\
I_{1c}&:=& \int_{0}^{1}ds\ll \iot \left( \tx_{0c}^{i*}(s)\,\dtx_{0c}^{i}(s)
-\dtx_{0c}^{i*}(s)\,\tx_{0c}^{i}(s)\right)\rr \nn \\
\partial &:=& -\iot \ll \tx_{0}^{i*''}\left( \tx_{0c}^{i}(1)-\tx_{0}^{i''}
\right) + \left( \tx_{0c}^{i*}(0)-\tx_{0}^{i*'}\right)\tx_{0}^{i'} \rr \; .\nn
\eea
Imposing periodic boundary conditions, (\ref{eq81}),(\ref{eq114}):
\be
\begin{array}{ccc}
\tx_{0}'=\tx_{0}''=:\tx_{0} &\hspace{1cm}& \e '=\e ''=:\e \\
\tx_{0}^{*'}=\tx_{0}^{*''}=:\tx_{0}^{*} &\hspace{1cm}& \e^{*'}=\e^{*''}
=:\e^{*}\; ,
\end{array}
\label{eqa15}
\ee
and using (\ref{eqa9}) and (\ref{eqa11}), one has:
\bea
I_{1c}&=&\int_{0}^{1}ds \ll \iot g_{0\l\g}\tp_{0}^{\l}\dot{\tp}_{c1}^{\g}(s)\rr
=-\e^{a*}\ll e^{i(\tcuf +\a\In)} \tcuf \rr^{ab} \e^{b} \nn \\
\partial &=& -\iot\ll \tx_{0}^{i*}\left( \tx_{0c}^{i}(1)-\tx_{0}^{i} \right)
+\left( \tx_{0}^{i*}(0)-\tx_{0}^{i*}\right) \tx_{0}^{i}\rr \; . \nn
\eea
Next one needs to use equations (\ref{eq36}),(\ref{eq38}), and (\ref{eqa8})
to compute $\tx_{0c}^{i}(1)$ and $\tx_{0c}^{i*}(0)$. In (\ref{eqa8}), the
boundary conditions must be chosen appropriately. The correct choice is
the following:
\[ \begin{array}{ccc}
\mbox{choose:}&\p (0)=\p_{0}=\frac{\tp_{0}}{\sqrt{\b}}\,\Rightarrow\,
\tp_{c1}(0)=0 & \mbox{ to compute:  $\tx_{0c}(1)$} \\
\mbox{choose:}&\p (\b )=\p_{0}=\frac{\tp_{0}}{\sqrt{\b}}\, \Rightarrow\,
\tp_{c1}(1)=0 & \mbox{ to compute:  $\tx_{0c}^{i*}(0)$}\; .
\end{array} \]
This leads to :
\[ \partial = \e^{a*}\ll e^{i(\tcuf +\a\In )}\tcuf \rr^{ab}\e^{b} \; ,\]
and hence
\be
I_{1}=0
\label{eqa16.9}
\ee

The computation of $I_{2}$ is straightforward. One replaces $\z$'s by
$\e_{0}$'s in (\ref{eqa4}), and sets $t'=0$ and $t''=1$. The final
result is obtained using (\ref{eqa6}),(\ref{eqa7}), and (\ref{eqa15}):
\be
I_{2}=-i\e^{a*}\ll e^{i(\tcuf +\a\In )} -\In \rr^{ab}\e^{b}\; .
\label{eqa17}
\ee
Combinning (\ref{eqa14}),(\ref{eqa16.9}), and (\ref{eqa17}), one has:
\[\so = -i\e^{a*}\ll e^{i(\tcuf +\a\In )}-\In \rr^{ab}\e^{b} \; .\]
This is used in Section~7, (\ref{eq119}). Taking $\tcuf = \e =\e^{*} =0$ ,
the situation reduces to the case of $\k =0$. In this case, one has
$\so = 0$.


\begin{thebibliography}{99}
\bibitem{bd1} B. S. DeWitt, {\em Supermanifolds}, Cambridge Univ.
Press, Cambridge (1992)
\bibitem{cd1} Y. Choquet-Bruhat, C. DeWitt-Morette, with M. Dillard-Bleick,
{\em Ananlysis, Manifolds and
Physics Parts I}, North-Holland, Amsterdam (1989); Y. Choquet-Bruhat and
C. DeWitt-Morette, {\em Analysis, Manifolds and Physics Part I\hspace{-.4mm}I},
North-Holland, Amsterdam (1989)
\bibitem{egh} T. Eguchi, P. B. Gilkey and A. J. Hanson:
``Gravitation, gauge theories and differential geometry,''
 Phys. Rep. {\bf 66} (1980) 213
\bibitem{witten} E. Witten: ``Constraints on supersymmetry breaking,''
Nucl. Phys. {\bf B202} (1982) 253
\bibitem{windey} P. Windey: ``Supersymmetric quantum mechanics and the
Atiyah-Singer index theorem,'' Acta Physica Polonica {\bf B15} No:5
(1984) 435
\bibitem{ag1} L. Alvarez-Gaum\'{e}: ``Supersymmetry and the Atiyah-Singer
index theorem,'' Commun. Math. Phys. {\bf 90}
(1983) 161
\bibitem{ag2} L. Alvarez-Gaum\'{e}: ``A note on the Atiyah-Singer index
theorem,'' J. Phys. {\bf A16} no:5 (1983) 4177
\bibitem{ag3} L. Alvarez-Gaum\'{e}, ``Supersymmetry and Index Theorem''
in {\em Supersymmetry}, Proceedings of the 1984 NATO School at Bonn,
Eds. K. Dietz, R. Flume, G. V. Gehlen and V. Rittenberg,  (1984)
\bibitem{manes-zumino} J. Ma\~{n}es and B. Zumino:
``WKB method, SUSY quantum mechanics and the index theorem,'' Nucl. Phys.
{\bf B270} (1986) 651
\bibitem{goodman} M. W. Goodman: ``Proof of character-valued index
theorems,'' Commun. Math. Phys. {\bf 107}, (1986) 391
\bibitem{getzler} E. Getzler: ``Pseudodifferential operators on
supermanifolds and the Atiyah-Singer index theorem,''
Commun. Math. Phys. {\bf 92} (1983) 163
\bibitem{friedan-windey} D. Friedan and P. Windey: ``Supersymmetric
derivation of the Atiayh-Singer index theorem and the chiral anomaly,''
Nucl. Phys. {\bf B235} (1984) 395
\bibitem{atiyah-singer} M. F. Atiyah and I. M. Singer, Bull. Am. Math.
Soc. {\bf 69} (1963) 422; M. F. Atiyah and I. M. Singer, Ann. Math.
{\bf 87} (1968) 484; ibid., 546
\bibitem{abp} M. F. Atiyah, R. Bott and V. K. Patodi: ``On the heat
equation and the index theorem,'' Inv. Math. {\bf 19} (1973) 279
\bibitem{atiyah} M. F. Atiyah, {\em The Index of Elliptic Operators},
Colloquium Lectures, Amer. Math. Soc., Dallas (1973), in Collected
works vol. III, Oxford Uni. Press (1988)
\bibitem{palias} R. S. Palais, {\em Seminar on the Atiyah-Singer Index
Theorem}, Ann. of Math. Study, vol. 57, Princeton Univ. Press (1965)
\bibitem{shanahan} P. Shanahan, {\em The Atiyah-Singer Index Theorem},
Lect. Notes in Math., vol. 638, Springer (1978)
\bibitem{gilkey} P. B. Gilkey, {\em Invariance Theory, the Heat Equation
and the Atiyah-Singer Index Theorem},  Math. Lecture series vol. 11,
Publish or Perish (1984)
\bibitem{booss} B. Booss and D. D. Bleecker, {\em Topology and Analysis},
Springer (1985)
\bibitem{berline} N. Berline, E. Getzler, M. Vergne, {\em Heat Kernels
and Dirac Operators}, Springer (1991)
\bibitem{atiyah-k} M. F. Atiyah, {\em K Theory}, W. A. Benjamin, Inc. (1967)
\bibitem{nash} C. Nash, {\em Differential Topology and Quantum Field
Theory}, Academic Press (1991)
\bibitem{susy} L. E. Gendenshte$\check{\imath}$n and I. V. Krive:
``Supersymmetry in quantum mechanics,'' Sov. Phys. Usp. {\bf 28}(8) (1986) 645
\bibitem{peierls} R. E. Peierls: ``The commutation laws of relativistic
field theory,'' Proc. Roy. Soc. (London) {\bf A214} (1952) 143
\bibitem{R-factor} A. Mostafazadeh: ``Supersymmetry and the Atiyah-Singer
Index Theorem I\hspace{-.4mm}I: The Scalar Curvature Factor in the
Schr\"{o}dinger Equation'' (1993)
\bibitem{cecotti} S. Cecotti and L. Girardello:
``Functional measure, topology and dynamical supersymmetry breaking,''
Phys. Lett. {\bf 110B}, no:1 (1982) 39
\bibitem{it-zu} C. Itzykson and J. B. Zuber, ``Quantum Field Theory'',
McGraw-Hills (1985)
\bibitem{faddeev} L. D. Faddeev, ``Introduction to Functional Methods''
in Les Houches Lectures XXVIII (1975), ``Methods in Field Theory'',
Ed.'s: R. Balian and J. Zinn-Justin, North-Holland.
\end{thebibliography}
\end{document}